\documentclass[a4paper,twocolumn,11pt]{quantumarticle}
\pdfoutput=1
\usepackage[utf8]{inputenc}
\usepackage[english]{babel}
\usepackage[T1]{fontenc}
\usepackage{hyperref}
\usepackage{tikz}
\usepackage{lipsum}
\usepackage{longtable}

\usepackage[numbers,sort&compress]{natbib}

\usepackage{amsmath,amssymb}
\usepackage{multirow}

\newcommand{\id}{\mathbb{I}}

\begin{document}

\title{Multi-Parameter Multi-Critical Metrology of the Dicke Model}

\author{Luca Previdi}
\affiliation{Dipartimento di Fisica e Astronomia, Università di Bologna, via Irnerio 46, I-40126 Bologna, Italy}
\email{lucaprevidi0202@gmail.com}
\orcid{0009-0008-6320-2079}
\author{Yilun Xu}
\affiliation{State Key Laboratory for Mesoscopic Physics, School of Physics, Frontiers Science Center for Nano-optoelectronics, Peking University, Beijing 100871, China}
\author{Qiongyi He}
\affiliation{State Key Laboratory for Mesoscopic Physics, School of Physics, Frontiers Science Center for Nano-optoelectronics, Peking University, Beijing 100871, China}
\affiliation{Collaborative Innovation Center of Extreme Optics, Shanxi University, Taiyuan, Shanxi 030006, China}
\affiliation{Hefei National Laboratory, Hefei 230088, China}
\author{Matteo G. A. Paris}
\affiliation{Quantum Technology Lab, Dipartimento di Fisica Aldo Pontremoli, Università degli Studi di Milano, I-20133 Milano, Italy}
\affiliation{INFN, Sezione di Milano, I-20133 Milano, Italy}

\maketitle

\begin{abstract}
 Critical quantum metrology exploits the hypersensitivity of quantum systems near phase transitions to achieve enhanced precision in parameter estimation. While single-parameter estimation near critical points is well established, the simultaneous estimation of multiple parameters, which is essential for practical sensing applications, remains challenging. This difficulty arises from \textit{sloppiness}, a phenomenon that typically renders the quantum Fisher information matrix (QFIM) singular or nearly singular. In this work, we demonstrate that multiparameter critical metrology is not only feasible but can also retain divergent precision scaling, provided one accepts a trade-off in the scaling exponent. Using the ground state of the single-cavity Dicke model (DM), we show that two Hamiltonian parameters can be simultaneously estimated with a scalar variance bound scaling as the square root of the critical parameter. This overcomes the inherent sloppiness by leveraging higher-order contributions to the QFIM. To recover the optimal quadratic scaling, we introduce the Dicke dimer (DD) with photon hopping. In this extended model, a triple point in the phase diagram enables the simultaneous closure of two excitation gaps, which effectively increases the rank of the QFIM and restores the ideal single-parameter scaling for specific parameter pairs. Furthermore, we extend our analysis to dissipative settings subject to photon loss. Finally, we establish a connection between the derived critical scalings and the fundamental state preparation time, providing a unified framework to operationally compare different sensing strategies. Our results demonstrate that critical quantum metrology can be made robust against dissipation and scalable to multiparameter scenarios, paving the way for practical quantum sensors operating near phase transitions.
\end{abstract}

\section{Introduction}
Critical metrology~\cite{garbe_critical_2020,mihailescu_critical_2025, zanardi_quantum_2008,montenegro2025quantum,fq4l-8v5g} is a rapidly developing field that exploits the extreme sensitivity of quantum systems near phase transitions to estimate physical parameters with enhanced precision. As a system approaches a quantum critical point, it becomes hypersensitive to microscopic perturbations, leading to a dramatic increase in the distinguishability between quantum states associated with infinitesimally different parameters. This susceptibility manifests as a diverging quantum Fisher information (QFI), the fundamental quantity that bounds the achievable precision in parameter estimation according to the quantum Cramér-Rao bound~\cite{paris_quantum_2009,liu_quantum_2020, giovannetti_advances_2011,chang_multiparameter_2025}. Specifically, for a control parameter $g$ driving the transition, the QFI for an estimated parameter typically diverges as $\sim |g_c - g|^{-2}$ at the leading order, where $g_c$ denotes the critical point~\cite{garbe_critical_2020, hotter_combining_2024, hotter_quantum_2025}.

However, for practical quantum sensing applications, it is often crucial to estimate multiple unknown parameters simultaneously, a task known as multiparameter quantum metrology~\cite{albarelli_perspective_2020,liu_quantum_2020,pezze_advances_2025}. This introduces significant theoretical challenges \cite{horodecki_five_2020}. Near criticality, sensitivity often becomes highly directional in the parameter manifold; the system responds strongly to changes along one specific axis but remains virtually blind to orthogonal variations \cite{gietka_understanding_2022}. This macroscopic low-dimensionality leads to a singular or severely ill-conditioned Quantum Fisher Information Matrix (QFIM), a phenomenon widely referred to as \textit{sloppiness}~\cite{he_scrambling_2025}. From a practical point of view, the determinant of the QFIM vanishes, implying that the full set of parameters cannot be estimated independently. Instead, only specific linear combinations remain accessible. Significant research effort is currently directed toward understanding and mitigating these rank-deficiency limitations~\cite{suzuki_nuisance_2020, suzuki_quantum_2020, mihailescu_uncertain_2025, zhu_quantum_2024}.

In this work, we demonstrate that there exists a fundamental trade-off between multiparameter accessibility and the scaling of precision near the critical point. Specifically, we show that while the leading-order QFIM is typically singular, higher-order contributions can restore invertibility, albeit with a modified scaling exponent. We first illustrate this mechanism using the ground state (GS) of the single-cavity Dicke model (DM)~\cite{dicke_coherence_1954,baumann_dicke_2010,klinder_dynamical_2015,luo_quantum_2025}. By simultaneously estimating the coupling strength $g$ and the cavity frequency $\omega_c$ near the critical point, we obtain an invertible QFIM where the scalar bound on the total variance vanishes as $\mathrm{Tr}[Q^{-1}] \sim |g_c - g|^{1/2}$. This represents a novel result: the simultaneous estimation of multiple parameters with vanishing error in a critical system. However, this method comes with two limitations: it is restricted to two parameters, and the convergence of the bound is slower than the ideal quadratic scaling characteristic of single-parameter estimation.

To overcome these limitations and recover faster scaling, we introduce the Dicke dimer (DD) model with photon hopping $\xi$~\cite{wei_boundary-induced_2025,xu_phase_2024}. This lattice model exhibits a triple point (TP) in the $\xi$--$g$ phase diagram where two distinct excitation gaps close simultaneously. Similar systems have recently been employed for quantum sensing tasks \cite{cheng_super-heisenberg_2025,mondal_multicritical_2025}. We argue that this additional source of criticality effectively increases the rank of the singular QFIM. By tuning the system's trajectory near this TP, we demonstrate that specific pairs of parameters can be estimated with a variance bound scaling of $\mathrm{Tr}[Q^{-1}] \sim |g_t - g|^2$, where $g_t$ denotes the value of $g$ at the triple point. Consequently, we are able to successfully recover the ideal single-parameter quadratic scaling within a multiparameter setting.

Furthermore, to address experimental realism, we analyze the steady state (SS) under photon loss, finding that the critical enhancement is not washed out by environmental noise. Indeed, the simultaneous estimation of multiple parameters remains possible with diverging precision, evidenced by a scalar variance bound that vanishes linearly as $C_S \sim |g_c-g|$. These results are particularly significant because they demonstrate that critical metrology can remain robust against dissipation \cite{zhao2025near}, preserving divergent precision even when employing a steady state as a probe.

Finally, we establish a connection between the critical scaling laws and the fundamental resource of time. By accounting for the inherent time overheads of these protocols—such as adiabatic preparation and steady-state relaxation times—we provide a rigorous operational framework to efficiently compare the true advantages of different estimation strategies.

Our findings pave the way toward practical implementations of critical quantum sensors, where multiple system parameters can be estimated simultaneously with high precision near phase transitions, making such protocols highly promising for real-world quantum sensing applications.

This work is structured as follows. In Section \ref{sec:QET_review}, we review the necessary tools from multiparameter quantum estimation theory, emphasizing the QFIM and the definition of sloppiness. Section \ref{sec:model} introduces the single-cavity Dicke model and the Dicke dimer, detailing their exact solutions in the thermodynamic limit and their steady states under photon loss. In Section \ref{sec:gs_estimation}, we present our main results on multiparameter estimation using the ground state, analyzing both the DM and the DD and exploiting TPs. Section \ref{sec:steady_state_estimation} extends this analysis to the open-system steady state, examining the robustness of the protocols against dissipation. Section \ref{sec:comparison} compares the different strategies and summarizes our findings. We conclude in Section \ref{sec:conclusions} with an outlook on future research directions.

\section{Multi-parameter Quantum Estimation Theory}
\label{sec:QET_review}

In this section, we briefly introduce the fundamental tools of quantum estimation theory employed in this work. For a comprehensive review, we refer the reader to Refs.~\cite{albarelli_perspective_2020,liu_quantum_2020,demkowicz-dobrzanski_multi-parameter_2020,pezze_advances_2025}.

\subsection{Matrix Cramér--Rao Bounds}
\label{sec:metrology}

\subsubsection{Classical Bound}
Consider a parametric family of states $\rho_{\boldsymbol{\lambda}}$ that depends on a vector of $d$ real parameters $\boldsymbol{\lambda} = (\lambda_1, \dots , \lambda_d)^T \in \mathbb{R}^d$.
To estimate these parameters, one must perform a measurement on the system, mathematically described by a positive operator-valued measure (POVM) $\hat{\Pi} = \{ \hat{\Pi}_k \, | \, \hat{\Pi}_k \ge 0,\, \sum_k \hat{\Pi}_k = \id \}$.
The probability $p(k|\boldsymbol{\lambda})$ of obtaining the outcome $k$ is given by the Born rule:
\begin{equation}
    p(k|\boldsymbol{\lambda}) = \text{Tr}[\rho_{\boldsymbol{\lambda} } \hat{\Pi}_k] \,.
\end{equation}
To infer the parameters from the measurement outcomes, we employ an estimator $\tilde{\boldsymbol{\lambda}}(k)$, which is a function mapping the measurement outcomes to the parameter space (e.g., the maximum likelihood estimator).
In the multiparameter setting, the estimation error is quantified by the mean square error covariance matrix:
\begin{equation}
    V(\boldsymbol{\lambda},\{ \hat{\Pi}_k\} ) = \sum_k p(k| \boldsymbol{\lambda} ) \big( \tilde{\boldsymbol{\lambda}}(k) -\boldsymbol{\lambda }\big) \big( \tilde{\boldsymbol{\lambda}}(k) -\boldsymbol{\lambda }\big)^T\,.
\end{equation}
Assuming the estimator is locally unbiased (i.e., unbiased in the neighborhood of the true parameter value), one has $\sum_k p(k| \boldsymbol{\lambda} ) \tilde{\boldsymbol{\lambda}}(k) = \boldsymbol{\lambda}$, and $V$ strictly corresponds to the covariance matrix (CM).

We then introduce the Fisher information matrix (FIM), whose elements are defined as:
\begin{align}\label{fim}
    F_{\mu \nu}(\boldsymbol{\lambda},\{\hat\Pi_k\} ) \equiv \sum_k \frac{\partial_{\mu} p(k|\boldsymbol{\lambda}) \partial_{\nu} p(k| \boldsymbol{\lambda})}{p(k| \boldsymbol{\lambda})}\,,
\end{align}
where $\partial_{\mu} \equiv \frac{\partial}{\partial \lambda_{\mu}}$. The FIM quantifies the amount of information about the parameters contained in the probability distribution $p(k|\boldsymbol{\lambda})$.
Under the local unbiasedness assumption, the covariance matrix is lower-bounded by the Cramér--Rao bound (CRB) \cite{cramer1999mathematical,Lehmann1998theor}:
\begin{equation}\label{CRB}
    V(\boldsymbol{\lambda},\{ \hat\Pi_k\} ) \ge \frac{1}{M}F(\boldsymbol{\lambda},\{\hat\Pi_k\} )^{-1}\,,
\end{equation}
where $M$ is the number of independent repetitions of the experiment. The inequality holds in the Loewner order sense, meaning the matrix difference $V - (MF)^{-1}$ is positive semidefinite.
In regular statistical models, Eq.~\eqref{CRB} represents an asymptotically achievable bound for consistent estimators (such as the Maximum Likelihood Estimator) in the limit $M\to\infty$.

\subsubsection{Quantum Bound}
Since the probability distribution depends on the choice of measurement $\hat{\Pi}$, it is natural to seek the measurement that maximizes the extracted information. While optimizing the FIM over all possible POVMs yields a unique solution for single-parameter estimation, the multiparameter case is nontrivial because different parameters may require incompatible optimal measurements.

However, it is possible to derive a measurement-independent upper bound on the FIM that depends solely on the quantum state family. The most prominent generalization is based on the symmetric logarithmic derivative (SLD) operators $\hat{L}_{\mu}$, defined implicitly by the Lyapunov equation:
\begin{align}\label{sldapp}
     \partial_{\mu } \rho_{\boldsymbol{\lambda}} &= \frac{ \hat{L}_{\mu} \rho_{\boldsymbol{\lambda}}+ \rho_{\boldsymbol{\lambda}} \hat{L}_{\mu} }{2}\,.
\end{align}
This leads to the definition of the QFIM~\cite{Helstrom1967,paris_quantum_2009,liu_quantum_2020}:
\begin{align}
    Q_{\mu \nu}(\boldsymbol{\lambda}) &\equiv \text{Tr} \bigg[\rho_{\boldsymbol{\lambda}}  \frac{\hat{L}_{\mu} \hat{L}_{\nu}+ \hat{L}_{\nu} \hat{L}_{\mu}}{2} \bigg] \label{QFIM} \,.
\end{align}
The QFIM provides the ultimate bound on precision, known as the quantum Cramér-Rao bound (QCRB):
\begin{align}\label{bound}
    V(\boldsymbol{\lambda},\{\hat\Pi_k\}) & \notag \ge \frac{1}{M}F(\boldsymbol{\lambda},\{\hat\Pi_k\})^{-1} \\ &\ge \frac{1}{M}Q(\boldsymbol{\lambda})^{-1}\,.
\end{align}
For pure-state models where $\rho_{\boldsymbol{\lambda}} = | \psi \rangle \langle \psi|$, the QFIM simplifies to:
\begin{align}\label{QFIpure}
     Q_{\mu \nu}(\boldsymbol{\lambda}) = & \notag 4 \text{Re} \big[\langle \partial_{\mu}\psi |\partial_{\nu} \psi \rangle \\&- \langle \partial_{\mu}\psi | \psi \rangle\langle \psi | \partial_{\nu}\psi \rangle \big].
\end{align}

\subsection{Scalar Cramér-Rao Bounds}
Since matrix optimization is only partially ordered, it is common to introduce a scalar figure of merit to unambiguously benchmark estimation efficiency. The standard choice is the weighted trace of the covariance matrix, $\text{Tr} \left[ W \, V(\boldsymbol{\lambda},\{\hat\Pi_k\} ) \right]$, where $W > 0$ is a positive definite weight matrix. Physically, this quantity represents a weighted sum of the variances of the estimators. For the simplest case of equal weighting, $W=\id_d$, it corresponds to the sum of the individual variances.

Applying this scalarization to the matrix bound in Eq.~\eqref{bound}, we obtain:
\begin{align}\label{eq:scalar_CRB}
     \text{Tr} \left[ W \, V(\boldsymbol{\lambda},\{\hat\Pi_k\} ) \right] &\ge \notag \frac{1}{M} \text{Tr} \left[F(\boldsymbol{\lambda},\{\hat\Pi_k\})^{-1} \right]\\ &\ge \frac{1}{M} C_S(\boldsymbol{\lambda}, W)\,,
\end{align}
where we have defined the SLD scalar variance bound:
\begin{align}
    C_S(\boldsymbol{\lambda}, W) &\equiv \text{Tr}[W Q^{-1}] \,.\label{scalar}
\end{align}
Throughout this work, we assume equal weighting for all parameters by setting $W = \id_d$. We will thus use the scalar quantity $C_S \equiv C_S(\boldsymbol{\lambda}, \id_d)$ to benchmark the sensing capabilities of the considered systems; the overall estimation precision is therefore related to $1/C_S$.

The scalar classical CRB, given by the first inequality of Eq.~\eqref{eq:scalar_CRB}, is asymptotically attainable in the limit $M \to \infty$ (i.e., for a large number of repetitions). Since an independent copy of the state must be prepared for each experimental repetition, the entire procedure consumes $M$ copies of the state, meaning the overall state is $\rho_{\boldsymbol{\lambda}}^{\otimes M}$. Crucially, in the quantum regime, it is theoretically possible to measure all the copies collectively rather than individually. Considering this collective measurement strategy, it is possible to prove the following inequality \cite{carollo2019quantumness,albarelli2020upper,candeloro2021properties,UHLMANN1986229}:
\begin{align}
    2 C_S(\boldsymbol{\lambda}, W) & \notag \ge M \min_{\{\hat{\Pi}_k\}}\text{Tr} \left[ W \, V(\boldsymbol{\lambda},\{\hat\Pi_k\} ) \right] \\ &\ge C_S(\boldsymbol{\lambda}, W) \,.
\end{align}
This implies that, even if the second inequality of Eq.~\eqref{eq:scalar_CRB} is not strictly tight, the minimum achievable variance $\text{Tr} \left[ W \, V(\boldsymbol{\lambda},\{\hat\Pi_k\} ) \right]$ remains upper-bounded by $2 C_S$. Most importantly for our analysis, if $C_S$ vanishes polynomially as the system approaches criticality, the true measurement variance must also vanish with the identical scaling behavior. This mathematical guarantee establishes the scalar variance bound $C_S$ as a robust and significant figure of merit for characterizing multiparameter sensitivity.

\subsubsection{Scalar Variance Bound vs. Sloppiness}
Crucially, the chain of matrix inequalities in Eq.~\eqref{bound} inherently requires the QFIM $Q$ to be invertible. A system for which $\det[Q] = 0$ is referred to as \textit{sloppy}. In such regimes, the simultaneous, independent estimation of the full parameter set is impossible \cite{he_scrambling_2025, suzuki_nuisance_2020}, because the system state effectively depends only on a reduced number of linear combinations of the parameters. To quantify this behavior, we define the sloppiness coefficient (SC) as \cite{he_scrambling_2025}:
\begin{equation}
    \mathcal{S} = \frac{1}{\det[Q]}.
\end{equation}

A common situation encountered in critical metrology—and central to this work—is a scenario in which the QFIM is singular at leading order but regains full rank when higher-order contributions are considered. To model this, let us assume the QFIM can be expanded in terms of a small parameter $\varepsilon \to 0$ (e.g., the distance to the critical point) as:
\begin{equation}
    Q = \varepsilon^{-k}(A+ \varepsilon B),
\end{equation}
where $A, B = \mathcal{O}(1)$. Here, we assume the leading-order matrix $A$ is singular, but the perturbed matrix $A+ \varepsilon B$ is full rank, with the overall scaling exponent typically being $k>0$. Under these conditions, the sloppiness coefficient scales as:
\begin{equation}
    \mathcal{S} \sim \frac{\varepsilon^{k-1}}{\sum_{i = 1}^d \det[A^{(i),B}]},
\end{equation}
where $A^{(i),B}$ denotes the matrix $A$ with its $i$-th row and column replaced by the $i$-th row and column of $B$. Correspondingly, the scalar variance bound scales as:
\begin{equation}
    C_S = \varepsilon^{k-1} \text{Tr}[(PBP)^{+}],
\end{equation}
where $(\cdot)^+$ denotes the Moore-Penrose pseudoinverse and $P = \id - A^{+} A$ is the projector onto the kernel of $A$. 

This mathematical structure reveals that the asymptotic scaling in $\varepsilon$ of the sloppiness coefficient $\mathcal{S}$ and the scalar bound $C_S$ are intrinsically linked. Furthermore, it demonstrates a vital concept: even if the leading-order QFIM is singular (rendering the exact critical point $\varepsilon=0$ non-invertible), one can still obtain a vanishing (divergently precise) scalar variance bound by carefully evaluating the higher-order contributions in the expansion.

\section{The Models}
\label{sec:model}

In this work, we study two paradigmatic models of light--matter interaction that exhibit quantum critical behavior: the single-cavity Dicke model and the Dicke dimer with photon hopping. Both models are analytically tractable in the thermodynamic limit and provide a rich platform for exploring multiparameter critical metrology.
\begin{figure}[t]
    \centering
    \includegraphics[width=0.9\linewidth]{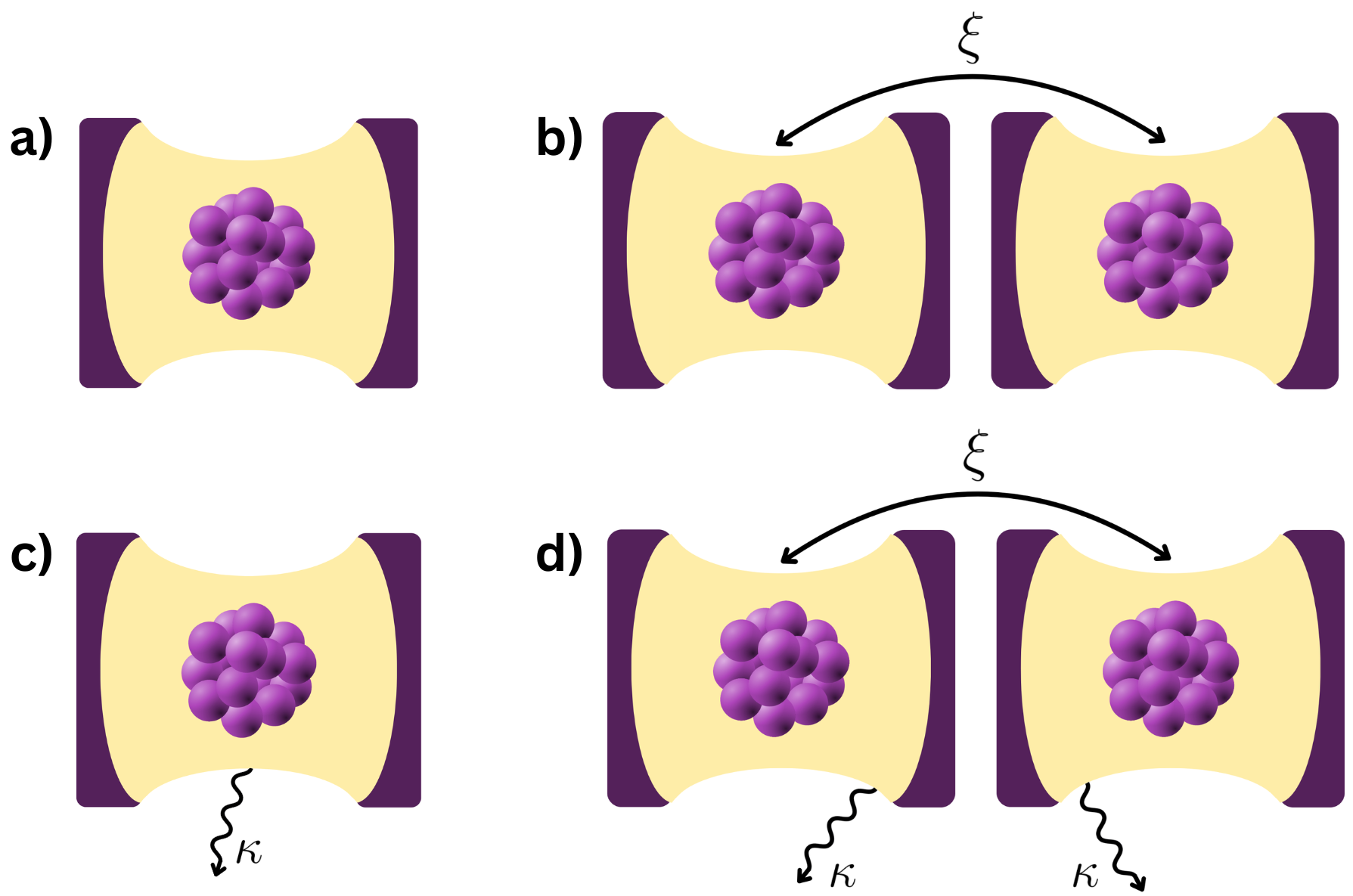}
    \caption{Pictorial representation of the cavity model considered. Namely: (a) single cavity Dicke model, (b) Dicke dimer, (c) single cavity Dicke model subject to photon loss, (d) Dicke dimer subject to photon loss.}
    \label{fig:rep_dicke}
\end{figure}

\subsection{Hamiltonians and Thermodynamic Limit}

The DM describes a single electromagnetic mode interacting with a collective atomic ensemble, pictorially represented in Fig. \ref{fig:rep_dicke}(a). Its Hamiltonian is given by
\begin{equation}
    H_{\text{DM}} = \omega_c \hat{a}^\dag \hat{a} + \omega_a \hat{S}^z + \frac{2g}{\sqrt{N_a}} (\hat{a} + \hat{a}^\dag) \hat{S}^x ,
    \label{eq:H_DM_original}
\end{equation}
where $\omega_c$ is the cavity frequency, $\hat{a}^\dag$ ($\hat{a}$) is the photonic creation (annihilation) operator, $\omega_a$ is the atomic transition frequency, $g$ is the light--matter coupling strength, and $N_a$ is the number of atoms. The collective spin operators are defined as $\hat{S}^j = \frac{1}{2} \sum_{k=1}^{N_a} \hat{\sigma}_{(k)}^j$ (with $j = x, y, z$), where $\hat{\sigma}_{(k)}^j$ are the Pauli matrices for the $k$-th atom.

The DD extends the above setting to two coupled cavities, each containing its own atomic ensemble, pictorially represented in Fig. \ref{fig:rep_dicke}(b). The Hamiltonian reads
\begin{align}
    H_{\text{DD}} =& \notag \sum_{n=1}^2 \left[ \omega_c \hat{a}_n^\dag \hat{a}_n + \omega_a \hat{S}_n^z + \frac{2g}{\sqrt{N_a}} (\hat{a}_n + \hat{a}_n^\dag) \hat{S}_n^x \right] \\ &+ \xi \left( \hat{a}_1^\dag \hat{a}_2 + \hat{a}_1 \hat{a}_2^\dag \right),
    \label{eq:H_DD_original}
\end{align}
where $\hat{a}_n^\dag, \hat{a}_n$ are the photonic operators for cavity $n=1,2$, $\hat{S}_n^j = \frac{1}{2} \sum_{k=1}^{N_a} \hat{\sigma}_{n,(k)}^j$ are the collective spin operators of the atoms in cavity $n$, and $\xi$ denotes the photon-hopping amplitude between the two cavities.

In the thermodynamic limit $N_a \to \infty$, the collective spin operators can be mapped to bosonic degrees of freedom via the Holstein--Primakoff transformation \cite{holstein_field_1940}:
\begin{align}
    &\hat{S}_n^+ = \sqrt{N_a - \hat{b}_n^\dag \hat{b}_n} \, \hat{b}_n, \\ &
    \hat{S}_n^- = \hat{b}_n^\dag \sqrt{N_a - \hat{b}_n^\dag \hat{b}_n}, \\ &
    \hat{S}_n^z = \frac{N_a}{2} - \hat{b}_n^\dag \hat{b}_n,
\end{align}
where $\hat{b}_n^\dag, \hat{b}_n$ are auxiliary bosonic operators. Expanding to leading order in $1/\sqrt{N_a}$ yields the effective bosonic Hamiltonians
\begin{align}
    H_{\text{DM}} =& \omega_c \hat{a}^\dag \hat{a} + \omega_a \hat{b}^\dag \hat{b} + g (\hat{a} + \hat{a}^\dag)(\hat{b} + \hat{b}^\dag), \label{eq:H_DM_bos} \\
     \notag H_{\text{DD}} =& \sum_{n=1}^2 \Big[ \omega_c \hat{a}_n^\dag \hat{a}_n + \omega_a \hat{b}_n^\dag \hat{b}_n \\ & \notag+ g (\hat{a}_n + \hat{a}_n^\dag)(\hat{b}_n + \hat{b}_n^\dag) \Big] 
     \\& + \xi \left( \hat{a}_1^\dag \hat{a}_2 + \hat{a}_1 \hat{a}_2^\dag \right). \label{eq:H_DD_bos}
\end{align}

For the DD, a further Bogoliubov transformation decouples the two cavities in the normal phase. Defining new normal-mode bosonic operators
\begin{align}
    \hat{A}_1 &= \frac{1}{\sqrt{2}} (\hat{a}_1 + \hat{a}_2), & \hat{A}_2 &= \frac{1}{\sqrt{2}} (\hat{a}_1 - \hat{a}_2), \\
    \hat{B}_1 &= \frac{1}{\sqrt{2}} (\hat{b}_1 + \hat{b}_2), & \hat{B}_2 &= \frac{1}{\sqrt{2}} (\hat{b}_1 - \hat{b}_2),
\end{align}
the Hamiltonian in Eq.~\eqref{eq:H_DD_bos} becomes
\begin{align}
    H_{\text{DD}} = & \notag \sum_{n=1}^2 \bigg[ \overline{\omega}_{c,n} \hat{A}_n^\dag \hat{A}_n + \omega_a \hat{B}_n^\dag \hat{B}_n \\ & + g (\hat{A}_n + \hat{A}_n^\dag)(\hat{B}_n + \hat{B}_n^\dag) \bigg],
\end{align}
where $\overline{\omega}_{c,n} = \omega_c + (-1)^n 2\xi$. Thus, the dimer decomposes into two independent single-cavity Dicke Hamiltonians with renormalized cavity frequencies $\overline{\omega}_{c,n}$.

\subsection{Ground States}
The ground state of the single-cavity Dicke model in the normal phase is a two-mode squeezed vacuum state \cite{zhou_quantum_2020}. It can be written as
\begin{equation}
    | G_{\text{DM}} \rangle = \hat{U} \hat{S}_a \hat{S}_b |0\rangle_a |0\rangle_b,
\end{equation}
where $|0\rangle_a, |0\rangle_b$ are the photonic and atomic vacuum states, respectively. The unitary $\hat{U}$ and squeezing operators $\hat{S}_a, \hat{S}_b$ are given by
\begin{align}
    \hat{U}  \notag=& \exp\bigg[ \frac{\theta}{2} \frac{\omega_a - \omega_c}{\sqrt{\omega_c \omega_a}} (\hat{a}^\dag \hat{b}^\dag - \hat{a} \hat{b}) 
    \\ &+ \frac{\theta}{2} \frac{\omega_a + \omega_c}{\sqrt{\omega_c \omega_a}} (\hat{a} \hat{b}^\dag - \hat{a}^\dag \hat{b}) \bigg], \\
    \hat{S}_a =& \exp\left[ \frac{r_a}{2} (\hat{a}^2 - \hat{a}^{\dag 2}) \right], \\
    \hat{S}_b =& \exp\left[ \frac{r_b}{2} (\hat{b}^2 - \hat{b}^{\dag 2}) \right],
\end{align}
with the parameters
\begin{align}\label{eq:theta}
    &\tan(2\theta) = -4g \frac{\sqrt{\omega_c \omega_a}}{\omega_c^2 - \omega_a^2}, \\
    &r_a= \frac{1}{2} \ln\left( \frac{\omega_+}{\omega_c} \right), \quad
    r_b = \frac{1}{2} \ln\left( \frac{\omega_-}{\omega_a} \right), \\
    & \notag\omega_\pm^2 = \frac{1}{2}(\omega_c^2 + \omega_a^2) \\ & \qquad\pm \frac{1}{2} \sqrt{ (\omega_c^2 - \omega_a^2)^2 + 16 g^2 \omega_c \omega_a }.
\end{align}
The excited states are obtained by applying creation operators to the ground state:
\begin{equation} \label{eq:eigenbasis_dm}
    |\psi_{j,k} \rangle_{a,b} = \hat{U} \hat{S}_a \hat{S}_b |j\rangle_a |k\rangle_b.
\end{equation}
For the Dicke dimer, the ground state factorizes into a product of two independent two-mode squeezed states:
\begin{align}
\label{eq:gs_dd}
    | G_{\text{DD}} \rangle = & \notag \big( \hat{U}_1 \hat{S}_{A_1} \hat{S}_{B_1} |0\rangle_{A_1}|0\rangle_{B_1} \big) 
    \\ &\otimes \big( \hat{U}_2 \hat{S}_{A_2} \hat{S}_{B_2} |0\rangle_{A_2}|0\rangle_{B_2} \big),
\end{align}
where $|0\rangle_{A_n}, |0\rangle_{B_n}$ are vacuum states corresponding to the ladder operators $\hat{A}_n$ and $\hat{B}_n$ respectively. Moreover, the operators $\hat{U}_{n}, \hat{S}_{A_n}, \hat{S}_{B_n}$ are given by:
\begin{align}
    \hat{U}_n  \notag=& \exp\bigg[ \frac{\theta_n}{2} \frac{\omega_a - \overline{\omega}_{c,n}}{\sqrt{\overline{\omega}_{c,n} \omega_a}} (\hat{A}_n^\dag \hat{B}_n^\dag - \hat{A}_n \hat{B}_n) 
    \\ &+ \frac{\theta_n}{2} \frac{\omega_a + \overline{\omega}_{c,n}}{\sqrt{\overline{\omega}_{c,n} \omega_a}} (\hat{A}_n \hat{B}_n^\dag - \hat{A}_n^\dag \hat{B}_n) \bigg], \\
    \hat{S}_{A_n} =& \exp\left[ \frac{r_{A_n}}{2} (\hat{A}_n^2 - \hat{A}_n^{\dag 2}) \right], \\ 
    \hat{S}_{B_n} =& \exp\left[ \frac{r_{B_n}}{2} (\hat{B}_n^2 - \hat{B}_n^{\dag 2}) \right],
\end{align}
with the parameters
\begin{align}
    &\tan(2\theta_n) = -4g \frac{\sqrt{\overline{\omega}_{c,n} \omega_a}}{\overline{\omega}_{c,n}^2 - \omega_a^2}, \\
    & r_{A_n} = \frac{1}{2} \ln\left( \frac{\omega_+}{\overline{\omega}_{c,n}} \right), \quad
    r_{B_n} = \frac{1}{2} \ln\left( \frac{\omega_-}{\omega_a} \right), \\
    & \notag\omega_{\pm,n}^2 = \frac{1}{2}(\overline{\omega}_{c,n}^2 + \omega_a^2) \\ & \qquad \pm \frac{1}{2} \sqrt{ (\overline{\omega}_{c,n}^2 - \omega_a^2)^2 + 16 g^2 \overline{\omega}_{c,n} \omega_a }.
\end{align}

The quantum phase transition occurs when the first energy gap vanishes. For the DM, the critical value of the coupling $g$ is
\begin{equation}
    g_c = \frac{\sqrt{ \omega_c \omega_a}}{2}.
\end{equation}
For the DD we have:
\begin{equation}
    g_c = \min_{k = 1,2} \frac{\sqrt{ \overline{\omega}_{c,k} \omega_a}}{2}\,.
\end{equation}
Throughout this work we will assume $g<g_c$ in order to remain in the so-called normal phase.

\subsection{Steady States under Photon Loss}
\label{sec:steady_state}

In any realistic implementation, photonic losses, pictorially represented in Fig. \ref{fig:rep_dicke}(c)(d), play a crucial role. We model dissipation via a Lindblad master equation. For the single-cavity DM, the dynamics is governed by
\begin{align}
\label{eq:lindblad_dm}
    \frac{d\rho(t)}{dt} &=  \notag -i [H_{\text{DM}}, \rho(t)] 
    \\ &+ \kappa \left( 2 \hat{a} \rho(t) \hat{a}^\dag - \hat{a}^\dag \hat{a} \rho(t) - \rho(t) \hat{a}^\dag \hat{a} \right), 
\end{align}
where $\kappa$ is the photon-loss rate. For the DD, we have
\begin{align}
\label{eq:lindblad_dd}
   & \frac{d\rho(t)}{dt} \notag = -i [H_{\text{DD}}, \rho(t)] 
    \\ &+ \kappa \sum_{n=1}^2 \left( 2 \hat{a}_n \rho(t) \hat{a}_n^\dag - \hat{a}_n^\dag \hat{a}_n \rho(t) - \rho(t) \hat{a}_n^\dag \hat{a}_n \right). 
\end{align}
Photonic losses shift the critical coupling. For the DM, the critical value becomes
\begin{equation}
    g_c = \frac{1}{2} \sqrt{ \omega_a \omega_c \left( 1 + \frac{\kappa^2}{\omega_c^2} \right) },
\end{equation}
while for the DD it is given by
\begin{equation}
    g_c = \min_{k=1,2} \frac{1}{2} \sqrt{ \omega_a \overline{\omega}_{c,k} \left( 1 + \frac{\kappa^2}{\overline{\omega}_{c,k}^2} \right) }.
\end{equation}
As for the GS, also in the noisy scenario we will assume $g<g_c$ in order to remain in the so-called normal phase.

We are interested in the SS $\rho_{\text{ss}} = \lim_{t\to\infty} \rho(t)$, satisfying $d\rho_{\text{ss}}/dt = 0$. Since the Hamiltonians are quadratic in the field operators and the Lindblad jump operators are linear, the dynamical map preserves Gaussianity \cite{weedbrook_gaussian_2012, ferraro_gaussian_2005, serafini_quantum_2021}. Consequently, given that the initial vacuum state is Gaussian, the resulting SS is guaranteed to be a continuous-variable Gaussian state.

Such states are fully characterized by their first and second moments. We introduce the vector of quadrature operators $\hat{\mathbf{u}} = (\hat{q}_a, \hat{p}_a, \hat{q}_b, \hat{p}_b)^\top$, with definitions $\hat{q}_j = (\hat{o}_j + \hat{o}_j^\dag)/\sqrt{2}$ and $\hat{p}_j = i(\hat{o}_j^\dag - \hat{o}_j)/\sqrt{2}$ (where $\hat{o} = \hat{a}, \hat{b}$). The state is then uniquely determined by the first-moment vector $\boldsymbol{d}$ and the covariance matrix $\boldsymbol{\sigma}$, with elements given by:
\begin{align}
    d_j &= \langle \hat{u}_j \rangle, \\
    \sigma_{jk} &= \frac{1}{2} \langle \{ \Delta \hat{u}_j, \Delta \hat{u}_k \} \rangle,
\end{align}
where $\Delta \hat{u}_j = \hat{u}_j - \langle \hat{u}_j \rangle$ and $\{ \cdot, \cdot \}$ denotes the anticommutator.

For the DM, the equations of motion for these moments are given by \cite{genoni_conditional_2016}
\begin{align}
    \label{eq:lyap_dm}\frac{d\boldsymbol{d}_{\text{DM}}}{dt} &= A_{\text{DM}} \boldsymbol{d}_{\text{DM}}, \\
    \frac{d\boldsymbol{\sigma}_{\text{DM}}}{dt} &= A_{\text{DM}} \boldsymbol{\sigma}_{\text{DM}} + \boldsymbol{\sigma}_{\text{DM}} A_{\text{DM}}^T + D_{\text{DM}},
    \label{eq:lyap_dd}
\end{align}
where $A_{\text{DM}}$ is the drift matrix and $D_{\text{DM}}$ is the diffusion matrix, given explicitly by
\begin{align}
\label{eq:drift_matrix_gs}
    A_{\text{DM}} =& \begin{pmatrix}
        -\kappa & \omega_c & 0 & 0 \\
        -\omega_c & -\kappa & -2g & 0 \\
        0 & 0 & 0 & \omega_a \\
        -2g & 0 & -\omega_a & 0
    \end{pmatrix}, \\ 
    D_{\text{DM}} =& \begin{pmatrix}
        2\kappa & 0 & 0 & 0 \\
        0 & 2\kappa & 0 & 0 \\
        0 & 0 & 0 & 0 \\
        0 & 0 & 0 & 0
    \end{pmatrix}.
\end{align}
In the SS, the first moments vanish ($\boldsymbol{d}_{\text{DM,ss}} = 0$) due to the form of $A_{\text{DM}}$. The steady-state covariance matrix is then obtained by solving the continuous Lyapunov equation:
\begin{equation}
    A_{\text{DM}} \boldsymbol{\sigma}_{\text{DM,ss}} + \boldsymbol{\sigma}_{\text{DM,ss}} A_{\text{DM}}^T + D_{\text{DM}} = 0. \label{eq:lyap_DM}
\end{equation}

For the DD, the formalism is identical. The SS is characterized by
\begin{align}
    &\boldsymbol{d}_{\text{DD,ss}}  = 0\,, \\
    &A_{\text{DD}} \boldsymbol{\sigma}_{\text{DD,ss}} + \boldsymbol{\sigma}_{\text{DD,ss}} A_{\text{DD}}^T + D_{\text{DD}}  = 0\,,
\end{align}
where the matrices decompose as $D_{\text{DD}} = D_{\text{DM}} \oplus D_{\text{DM}}$ and $A_{\text{DD}} = A_1 \oplus A_2$, with the block matrices
\begin{equation}
\label{eq:drift_matrix_ss}
    A_n = \begin{pmatrix}
        -\kappa & \overline{\omega}_{c,n} & 0 & 0 \\
        -\overline{\omega}_{c,n} & -\kappa & -2g & 0 \\
        0 & 0 & 0 & \omega_a \\
        -2g & 0 & -\omega_a & 0
    \end{pmatrix}\,,
\end{equation}
with $n = 1,2$.
Solving these linear equations provides the covariance matrix $\boldsymbol{\sigma}_{\text{ss}}$, which completely specifies the Gaussian SS.
\section{Multi-Parameter Estimation using the Ground State}
\label{sec:gs_estimation}

The ground state of critical quantum systems has served as a primary resource in quantum metrology, largely due to its experimental accessibility~\cite{baumann_dicke_2010}. Near a quantum phase transition, the closing of the energy gap renders the ground state exceptionally sensitive to small perturbations in the Hamiltonian parameters~\cite{garbe_critical_2020, zanardi_quantum_2008}. This critical susceptibility manifests as a diverging QFI for individual parameters, providing a rigorous pathway to enhanced estimation precision. Accordingly, we begin our analysis by establishing the multi-parameter precision limits inherent to the ground states of the single-cavity DM and the DD.

We will focus primarily on benchmarking the precision scaling near the critical point. This focus is motivated by the phenomenon of critical slowing down: as the energy gap closes, the time required to adiabatically prepare the ground state diverges. Consequently, the trade-off between the diverging precision and the diverging preparation time makes the asymptotic scaling behavior near $g_c$ the decisive factor for practical metrology.

Throughout this work, we define the vector of physical parameters to be estimated as $\boldsymbol{\lambda} = \{ \omega_c, g, \omega_a, \xi, \kappa \}$. We assign indices to these parameters as follows: $\lambda_1 = \omega_c$, $\lambda_2 = g$, $\lambda_3 = \omega_a$, $\lambda_4 = \xi$, and $\lambda_5 = \kappa$. We denote the QFIM relative to a specific subset of parameter indices $\{i_1, i_2, \dots, i_n\}$ as $Q^{\{i_1, i_2, \dots, i_n\}}$. For example, $Q^{\{1,3,5\}}$ refers to the QFIM calculated with respect to the subset $\{ \omega_c, \omega_a, \kappa \}$.

\subsection{Estimation with a Single Cavity}
\label{sec:single_cavity_gs}

In this section, we analyze the ground state of the DM, for which the parameters of interest are $\lambda_1 = \omega_c,\lambda_2 = g, \lambda_3 = \omega_a $. As discussed in Appendix \ref{sec_qfim_dm}, the elements of the QFIM relative to the ground state of the DM can be evaluated as:
\begin{align}
    Q_{\mu \nu}& \notag = 2(\partial_\mu r_{a}+\chi^\mu_{ab})(\partial_\nu r_{a}+\chi^\nu_{ab})\\ & \notag + 2(\partial_\mu r_{b}-\chi^\mu_{ab})(\partial_\nu r_{b}-\chi^\nu_{ab}) \\
    & \notag + (\cosh( r_{b}-r_{a}) \chi^\mu_{-}+ \sinh( r_{b}-r_{a}) \chi^\mu_{+}) \\
    &  (\cosh( r_{b}-r_{a}) \chi^\nu_{-}+ \sinh( r_{b}-r_{a}) \chi^\nu_{+})
\label{eq:QFIM_full}
\end{align}
where we have defined the auxiliary quantities:
\begin{align}
    c_\pm &= \theta \frac{\omega_a \pm \omega_c} { \sqrt{\omega_c \omega_a}}, \quad
    \delta_\mu = c_+ \partial_\mu c_- - \partial_\mu c_+ c_- \\
    \chi^\mu_{ab} &=\frac{\delta_\mu}{4 \theta^2}(1-4 \theta^2 -\cos(2 \theta)) \\
    \chi^\mu_{\pm} &=\partial_\mu c_\pm + \frac{\delta_\mu c_\mp}{8 \theta^3} ( 2 \theta- \sin(2 \theta)).
\end{align}
Near criticality ($g \to g_c$), the leading order (LO) term of the QFIM scales as:
\begin{align}
    Q_{\mu \nu} \approx 2 \partial_\mu r_b \partial_\nu r_b & \notag\approx \frac{1}{2 \omega_-^2}\partial_\mu \omega_-\partial_\nu \omega_-\\ & \equiv Q^{\text{LO}}_{\mu \nu} \sim \frac{1}{(g_c-g)^2}.
\end{align}
This matrix is clearly singular as it is of rank 1, meaning the system is sloppy and simultaneous estimation of two or more parameters is not feasible, at least at LO.

Indeed, while the leading order of the QFIM is generally non-invertible near criticality due to renormalization group constraints \cite{mihailescu_multiparameter_2024}, higher-order contributions can restore invertibility. Specifically, we consider the next-to-leading order terms:
\begin{align}
    (\cosh( r_{b}-r_{a}) \chi^\mu_{-}+& \notag \sinh( r_{b}-r_{a}) \chi^\mu_{+}) \\ & \approx \frac{1}{2}\sqrt{\frac{\omega_a \omega_+}{ \omega_c} } \frac{\Delta\chi^\mu}{\sqrt{\omega_-}}
\end{align}
where $\Delta \chi^\mu = \chi^\mu_- - \chi^\mu_+$, given by:
\begin{equation}
    \Delta \chi^\mu = \partial_\mu \Delta c+ \frac{\delta_\mu \Delta c}{8 \theta^3} (2 \theta - \sin(2 \theta))
\end{equation}
with $\Delta c = c_- - c_+ = -2 \theta \sqrt{\frac{\omega_c}{\omega_a}}$. We then define the asymptotic behavior matrices:
\begin{align}
    B^{\{i_1, \dots, i_n\}}_{\mu \nu} &= \mathcal{B} \Delta \chi^\mu \Delta \chi^\nu \\
    A^{\{i_1, \dots, i_n\}}_{\mu \nu} &= \lim_{g \rightarrow g_c^-} Q^{\{i_1, \dots, i_n\}}_{\mu \nu} (g_c-g)^2
\end{align}
where the constant prefactor is $\mathcal{B} = \frac{\omega_a^2 (\omega_a^2+\omega_c^2)}{8(\omega_a \omega_c)^{7/4}}$ and $\mu, \nu \in \{i_1, \dots, i_n \}$. Focusing specifically on the simultaneous estimation of cavity frequency and coupling strength ($\omega_c, g$), the leading-order matrix is:
\begin{equation}
    A^{\{1,2\}} = \begin{pmatrix}
     \frac{1}{8}  & - \sqrt{\frac{\omega_a}{\omega_c}} \frac{1}{32 }\\
     - \sqrt{\frac{\omega_a}{\omega_c}} \frac{1}{32 } & \frac{\omega_a}{ \omega_c} \frac{1}{128}
    \end{pmatrix}.
\end{equation}
Calculating the trace of the inverse QFIM (the scalar bound on total variance), we find:
\begin{equation}
    \operatorname{Tr}[(Q^{\{1,2\}})^{-1}] = \mathcal{T}^{\{1,2\}} \sqrt{g_c-g} + O(g_c-g)
\end{equation}
where the prefactor $\mathcal{T}^{\{1,2\}}$ is given by:
\begin{align}\label{eq:prefactor}
    \mathcal{T}^{\{1,2\}} & \notag= \frac{A_{gg}+ A_{\omega_c \omega_c}}{(\sqrt{A_{\omega_c \omega_c} B_{gg}}- \sqrt{A_{gg} B_{\omega_c \omega_c}} )^2} \\ &= \frac{\omega_a+ 16 \omega_c}{ \mathcal{B}(\sqrt{\omega_a} | \Delta \chi^g| - 4 \sqrt{\omega_c}| \Delta \chi^{\omega_c}|)^2}.
\end{align}

\begin{figure}[t]
    \centering
    \includegraphics[width=0.9\linewidth]{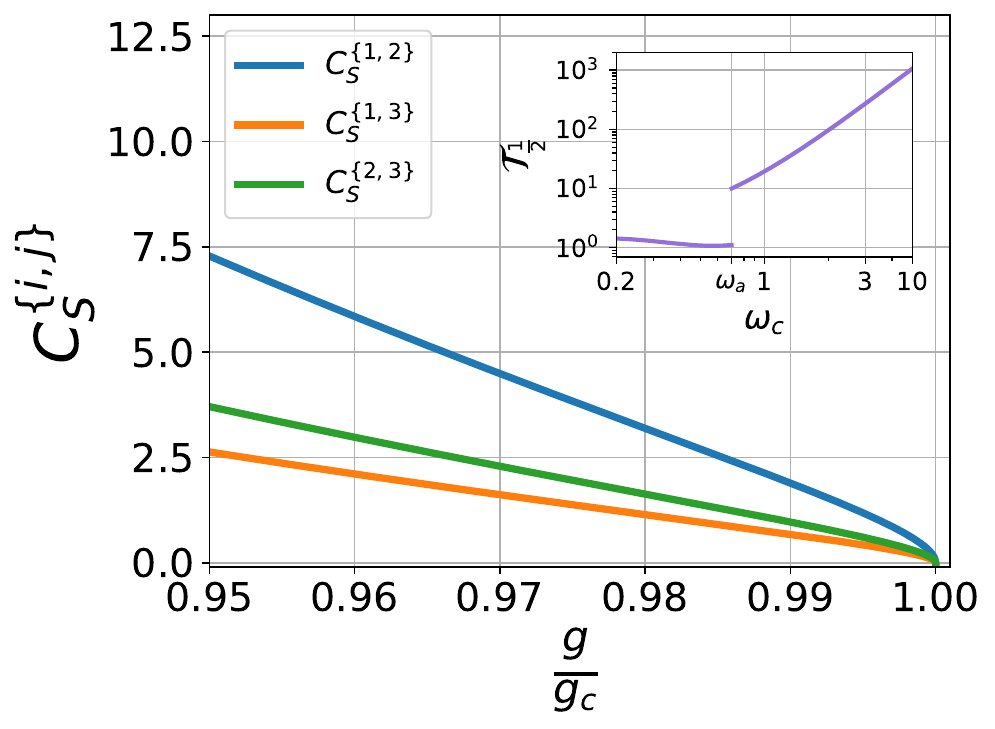}
    \caption{Plot of the scalar variance bound $C_S^{\{i,j\}} = \operatorname{Tr}[(Q^{\{i,j\}})^{-1}]$ for all pairwise combinations of parameters, as a function of the normalized coupling $g/g_c$. The simulation is performed employing the GS of the DM at fixed $\omega_c = 1$ and $\omega_a = 0.7$. The inset shows the prefactor $\mathcal{T}^{\{ 1,2\}}$, given by Eq.~\eqref{eq:prefactor}, as a function of $\omega_c$ for fixed $\omega_a = 0.7$.}
    \label{fig:DM_gs}
\end{figure}

The scalar variance bound $C_S^{\{i, j\}}$ is shown in Fig. \ref{fig:DM_gs} for all pairwise combinations of $i$ and $j$, explicitly showing the scaling $\sim \sqrt{g_c-g}$. In the inset of Fig. \ref{fig:DM_gs}, we plot the prefactor $\mathcal{T}^{\{1,2\}}$; a discontinuity is observed at $\omega_a = \omega_c$, arising from the definition of $\theta$ in Eq.~\eqref{eq:theta}, which is discontinuous at resonance.

Generally, the regime $\omega_c < \omega_a$ offers superior sensing performance, indicated by a lower prefactor $\mathcal{T}^{\{1,2\}}$. However, most experimental implementations operate in the regime $\omega_a < \omega_c$ \cite{baumann_dicke_2010, klinder_dynamical_2015}, which motivates the parameter choice in Fig. \ref{fig:DM_gs}.

A crucial trade-off is observed: while the scalar variance bound vanishes ($C_S^{\{i,j\}} \to 0$) as $g \to g_c$, verifying the possibility of infinitely precise estimation, the convergence rate $\sqrt{g_c-g}$ is significantly slower than the standard single-parameter scaling of $(g_c-g)^2$ \cite{garbe_critical_2020}.

Unfortunately, extending the analysis to three parameters makes the model sloppy again. Numerical simulations indicate that $\det[Q^{\{ 1,2,3\}}] = 0$ always; consequently, the simultaneous estimation of the triplet $\{ \omega_c, g, \omega_a \}$ is impossible, even at finite precision.

\subsection{Estimation with the Dicke Dimer}
\label{sec:DD_triple_point}

We now extend our analysis to the ground state of the DD, for which the parameters of interest are $\lambda_1 = \omega_c, \lambda_2 = g, \lambda_3 = \omega_a, \lambda_4 = \xi$. The model now allows for photonic hopping between the two cavities, quantified by the parameter $\xi$.

As derived in Appendix \ref{app:a2}, the general form of the QFIM for the DD is given by the sum of the contributions from the two decoupled normal modes:
{\footnotesize \begin{align}
    & \notag Q_{\mu \nu} = \sum_{n = 1}^2 \bigg[ 2(\partial_\mu r_{A_n}+\chi^\mu_{ab,n})(\partial_\nu r_{A_n}+\chi^\nu_{ab,n}) \\ & \notag + 2(\partial_\mu r_{B_n}-\chi^\mu_{ab,n})(\partial_\nu r_{B_n}-\chi^\nu_{ab,n}) \\
    & \notag   + (\cosh( r_{B_n}-r_{A_n}) \chi^\mu_{-,n}+ \sinh( r_{B_n}-r_{A_n}) \chi^\mu_{+,n})  \\
    &   \times (\cosh( r_{B_n}-r_{A_n}) \chi^\nu_{-,n}+ \sinh( r_{B_n}-r_{A_n}) \chi^\nu_{+,n}) \bigg]
\label{eq:QFIM_DD_sec}
\end{align}}
where we have defined the auxiliary quantities for each mode $n \in \{1,2\}$:
\begin{align}
    &c_{\pm,n} = \theta_n \frac{\omega_a \pm \overline{\omega}_{c,n}} { \sqrt{\overline{\omega}_{c,n} \omega_a}}, \\
    &\delta_{\mu,n} = c_{+,n} \partial_\mu c_{-,n} - \partial_\mu c_{+,n} c_{-,n}\,, \\
    &\chi^\mu_{ab,n} =\frac{\delta_{\mu,n}}{4 \theta_n^2}(1-4 \theta_n^2 -\cos(2 \theta_n))\,, \\
    &\chi^\mu_{\pm,n} =\partial_\mu c_{\pm,n} + \frac{\delta_{\mu,n} c_{\mp,n}}{8 \theta_n^3} ( 2 \theta_n- \sin(2 \theta_n)).
\end{align}

\begin{figure*}[t]
    \centering
    \includegraphics[width=0.9\linewidth]{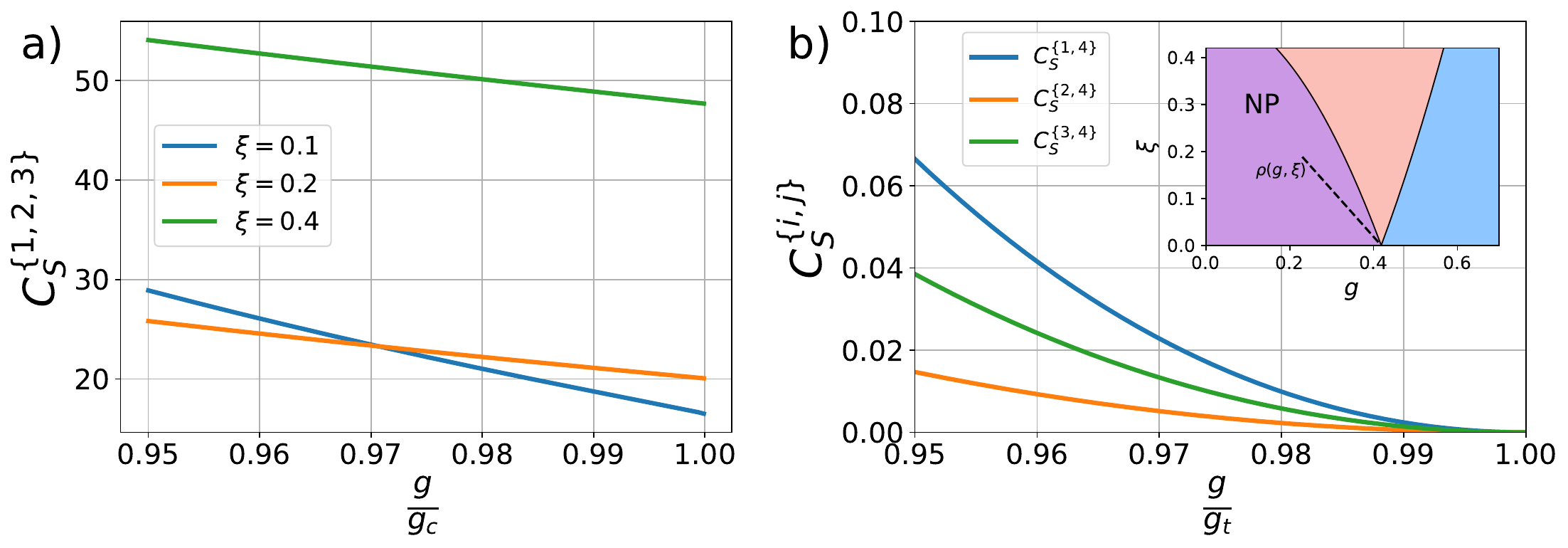}
    \caption{(a) Plot of the scalar variance bound $C_S^{\{1,2,3\}}$ for the triplet $\{\omega_c, g, \omega_a\}$ as a function of $g/g_c$ for fixed $\xi \in \{0.1, 0.2, 0.4\}$, employing the GS of the DD as a probe. (b) Plot of the scalar variance bound $C_S^{\{i,4\}} = \operatorname{Tr}[(Q^{\{i,4\}})^{-1}]$ for the simultaneous estimation of $\xi$ and one other parameter ($\omega_c, g,$ or $\omega_a$), as a function of $g/g_t$. The simulation has been performed employing the GS of DD as a probe and dynamically tuning both $g$ and $\xi$ toward the TP over a trajectory defined by Eq.~\eqref{eq:traj} with fixed $ k = 1$. The inset represents the phase diagram of the DD in the $g-\xi$ plane, with the trajectory highlighted. Both panels use fixed $\omega_c = 1, \omega_a = 0.7$.}
    \label{fig:DD_gs}
\end{figure*}

As shown in Fig. \ref{fig:DD_gs}(a), this structure allows for the estimation of the three parameters $\omega_c, g, \omega_a$ with finite (non-diverging) precision. Note that this was not possible in the single-cavity DM, where $\det[Q^{\{1,2,3\}}] = 0$ everywhere. However, there is no advantage in the scaling of the precision for two-parameter models. Indeed, for most of the pairs we find $C_S^{\{i,j\}} \sim (g_c-g)^{1/2}$, as in the single-cavity scenario, with the only exception of $C_S^{\{1,4\}}$ which does not vanish as $g \to g_c$. This implies that the simultaneous estimation of $\omega_c$ and $\xi$ is possible only at finite precision.

A potential solution to this scaling limitation lies in exploiting TPs. The core strategy involves tuning the system to the vicinity of a TP where two distinct energy gaps close simultaneously, introducing an additional source of criticality. Physically, we expect this to increase the rank of the diverging component of the QFIM, thereby mitigating the sloppiness observed in the single-cavity case and potentially enabling the simultaneous estimation of multiple parameters with optimal scaling.

A TP in the DD phase diagram occurs at the coordinates $(\xi_t, g_t) = (0, \frac{\sqrt{\omega_a \omega_c}}{2})$ in the $\xi-g$ plane. To exploit the critical properties of this point, we define a trajectory parametrized by $\varepsilon > 0$ that approaches the TP as $\varepsilon \to 0^+$:
\begin{equation} \label{eq:traj}
    g = g_t - \varepsilon, \quad \xi = k \varepsilon \,,
\end{equation}
where $k > 0$ is a proportionality constant. Near $\xi = 0$, the critical coupling line $g_c(\xi)$ decreases linearly with the photon hopping strength:
\begin{equation}
    g_c(\xi) = \frac{\sqrt{\omega_a \omega_c}}{2} - \frac{1}{2} \sqrt{\frac{\omega_a}{ \omega_c}}| \xi | + O (\xi^2).
\end{equation}
To ensure the trajectory remains within the normal phase as it approaches the TP, we require $k < k_{\text{max}} = 2 \sqrt{\omega_c / \omega_a}$ to stay below the critical line.

Crucially, at the TP, the eigenfrequencies of both lower polariton branches vanish: $\omega_{-,1}|_{(\xi_t, g_t)} = \omega_{-,2}|_{(\xi_t, g_t)} = 0$. Consequently, the Leading Order (LO) term of the QFIM is dominated by the divergence of both modes:
\begin{equation}
    Q_{\mu \nu} \approx Q^{\text{LO}}_{\mu \nu} = \sum_{n=1}^2 \frac{1}{2 \omega_{-,n}^2}\partial_\mu \omega_{-,n}\partial_\nu \omega_{-,n}.
\end{equation}
Unlike the single-cavity case, $Q^{\text{LO}}$ is now the sum of two rank-1 matrices. If the gradient vectors $\partial_\mu \omega_{-,1}$ and $\partial_\mu \omega_{-,2}$ are linearly independent, the rank of the LO matrix increases to 2.

Our analysis confirms that for the simultaneous estimation of the photon hopping $\xi$ and any other parameter from the set $\{\omega_c, g, \omega_a\}$, the matrix $Q^{\text{LO}, \{i,4\}}$ becomes invertible. As shown in Fig. \ref{fig:DD_gs}(b), this leads to a scaling of the scalar variance bound:
\begin{equation}
    \operatorname{Tr}[(Q^{\{i,4\}})^{-1}] \sim (g_c - g)^2.
\end{equation}
 Consequently, we recover the quadratic scaling characteristic of single-parameter critical metrology \cite{garbe_critical_2020}, but now for \textit{two} parameters simultaneously.

However, limitations remain when extending beyond two parameters. Indeed, while it is possible to invert the QFIM for the triplet $(\omega_c, g, \omega_a)$, the quantity $C_S^{\{1,2,3\}}$ saturates to a finite value rather than vanishing as $g \to g_c$, similar to the fixed $\xi$ case plotted in Fig. \ref{fig:DD_gs}(a). 


Regarding the estimation of parameter pairs that do \textit{not} include $\xi$ (e.g., $g$ and $\omega_c$), we observe no scaling advantage compared to the single-cavity DM. The precision scales as $C_S^{\{i,j\}} \sim \sqrt{g_c-g}$ whether tuning $\xi$ to a TP or fixing it a priori, consistent with the results in Section \ref{sec:single_cavity_gs}.

Finally, numerical simulations for the full four-parameter set yield $\det[Q^{\{1,2,3,4\}}] = 0$ everywhere, both when tuning $\xi$ and when fixing it. This confirms that, despite the additional structure provided by the TP, the model remains sloppy for the full parameter space, rendering the simultaneous estimation of all four parameters impossible even at finite precision.

\section{Multi-Parameter Estimation using the Steady State under Photon Loss}
\label{sec:steady_state_estimation}

We now extend our analysis to a more realistic scenario by incorporating photonic loss. The open-system dynamics are governed by the Lindblad master equation, as described in Eq.~\eqref{eq:lindblad_dm} for the DM and Eq.~\eqref{eq:lindblad_dd} for the DD. As discussed in Section \ref{sec:steady_state}, since the Hamiltonian is quadratic and the jump operators are linear, the SS is guaranteed to remain Gaussian. Furthermore, since the drift matrices from Eq.s~\eqref{eq:drift_matrix_gs} and \eqref{eq:drift_matrix_ss} are full rank, the first moments vanish in the SS ($\mathbf{d}_{ss} = 0$).

For a generic $n$-mode Gaussian state with vanishing first moments and covariance matrix $\boldsymbol{\sigma}$, the elements of the QFIM can be evaluated analytically using the vectorization formalism \cite{chang_multiparameter_2025}:
\begin{equation}
    Q_{\mu \nu } = \frac{1}{2} \text{vec} \left[ \frac{\partial \boldsymbol{\sigma} }{ \partial \lambda_\mu} \right]^T \left( \boldsymbol{\sigma} \otimes \boldsymbol{ \sigma}- \Omega \otimes \Omega\right)^{-1} \text{vec} \left[ \frac{\partial \boldsymbol{\sigma} }{ \partial \lambda_\nu} \right]
\end{equation}
where the vectorization operation $\text{vec}[M]$ transforms a matrix $M$ into a column vector by stacking its columns. For example, for a $2 \times 2$ matrix:
\begin{equation}
    M = \begin{pmatrix}
        a & b\\
        c & d
        \end{pmatrix}
        \quad \Longrightarrow \quad 
    \text{vec}[M] = \begin{pmatrix}
        a \\ c\\
        b \\ d
        \end{pmatrix}\,.
\end{equation}
The symplectic form $\Omega$ for an $n$-mode system is defined as the direct sum of $n$ single-mode symplectic matrices:
\begin{equation}
    \Omega = \bigoplus_{j =1}^n \Omega_1\,, \quad \text{with} \quad \Omega_1 = \begin{pmatrix}
        0 & 1\\
        -1 & 0
        \end{pmatrix}\,.
\end{equation}

Since $\boldsymbol{\sigma}$ represents the covariance matrix of a SS, it satisfies the continuous Lyapunov equation:
\begin{equation}
\label{eq:lyap_cm}
    A \boldsymbol{\sigma} + \boldsymbol{\sigma} A^T + D = 0,
\end{equation}
as detailed in Eqs.~\eqref{eq:lyap_dm} and \eqref{eq:lyap_dd}. To evaluate the QFIM, we require the derivatives of the covariance matrix with respect to the parameters, $\partial \boldsymbol{\sigma} / \partial \lambda_\mu$. Differentiating Eq.~\eqref{eq:lyap_cm} with respect to $\lambda_\mu$ yields a new Lyapunov equation for the derivative:
\begin{equation}
\label{eq:lyap_der_cm}
    A \frac{\partial \boldsymbol{\sigma}}{ \partial \lambda_\mu} + \frac{\partial \boldsymbol{\sigma}}{ \partial \lambda_\mu} A^T + \left( \frac{\partial A }{ \partial \lambda_\mu} \boldsymbol{\sigma} + \boldsymbol{\sigma} \frac{\partial A^T }{ \partial \lambda_\mu} + \frac{\partial D }{ \partial \lambda_\mu} \right) = 0\,.
\end{equation}
Eqs.~\eqref{eq:lyap_cm} and \eqref{eq:lyap_der_cm} form a closed system of linear algebraic equations that can be solved numerically with high efficiency. Crucially, this analytical approach allows us to determine $\partial \boldsymbol{\sigma} / \partial \lambda_\mu$ exactly, without relying on finite-difference methods. This is particularly important near critical points, where the divergent susceptibility can cause severe numerical instability in finite-difference schemes.

\subsection{Estimation with a Single Cavity}
\label{sec:single_cavity_ss}

In the single-cavity DM subject to photonic loss, we consider the parameter set $\lambda_1 = \omega_c, \lambda_2 = g, \lambda_3 = \omega_a, \lambda_5 = \kappa$.

\begin{figure}[t]
    \centering
    \includegraphics[width=0.9\linewidth]{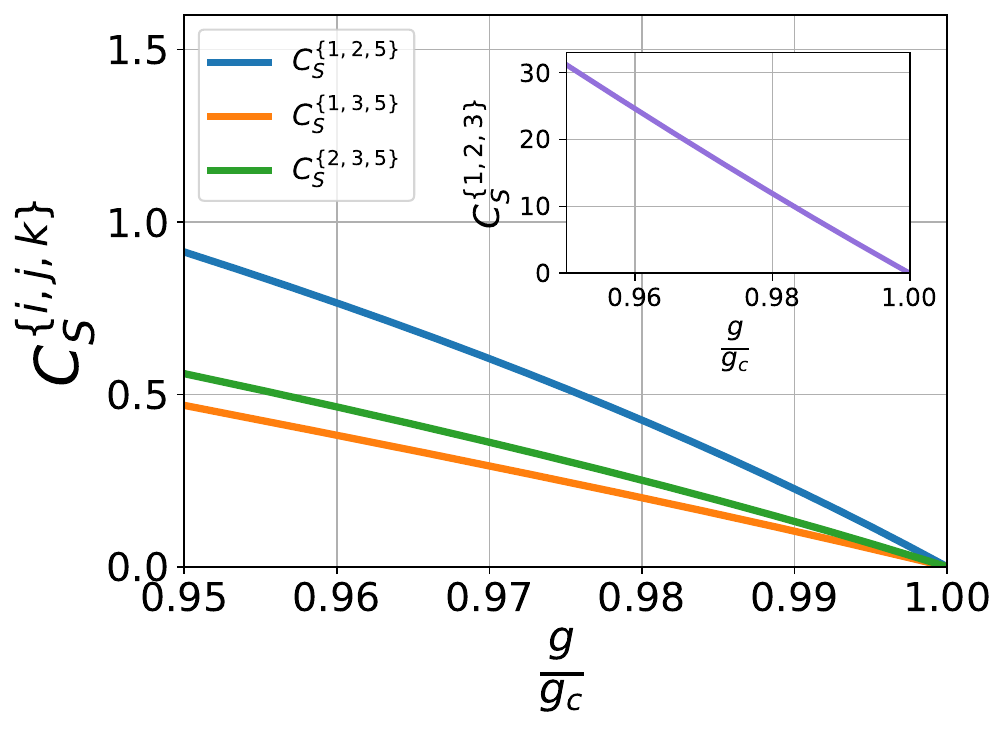}
    \caption{Plot of the scalar variance bound $C_S^{\{i,j,k\}}$ for the SS of the DM, as a function of the normalized coupling $g/g_c$. We fixed $\omega_a = 0.7$, $\omega_c = 1$ and $\kappa = 0.1$. The main panel shows the estimation of the decay rate $\kappa$ simultaneously with two other parameters. The inset panel shows the simultaneous estimation of the Hamiltonian parameters $\omega_c, g, \omega_a$.}
    \label{fig:DM_ss}
\end{figure}

Our results, shown in Fig.~\ref{fig:DM_ss}, indicate that it is possible to estimate subsets of up to three parameters simultaneously with a variance bound that vanishes linearly. Specifically, the scalar bound $C_S$ scales as:
\begin{equation}
    C_S^{\{i,j,k\}} \sim (g_c - g).
\end{equation}
This linear scaling implies that the estimation precision diverges as $(g_c-g)^{-1}$ as the system approaches the critical point. While this divergence is slower than the ideal quadratic scaling, it demonstrates a significant robustness against dissipation, which typically saturates precision in non-critical systems.

However, the "sloppiness" issue persists for the full parameter set. When extending the analysis to all four parameters simultaneously, we find that $\det[Q^{\{1,2,3,5\}}] = 0$ everywhere. Consequently, the simultaneous estimation of the full set $\{\omega_c, \omega_a, g, \kappa\}$ remains impossible, even at finite precision.

\subsection{Estimation with the Dicke Dimer}
\label{sec:DD_steady_state_TP}

In the DD under photonic loss, the set of unknown parameters expands to $\lambda_1 = \omega_c, \lambda_2 = g, \lambda_3 = \omega_a, \lambda_4 = \xi, \lambda_5 = \kappa$.

\begin{figure}
    \centering
    \includegraphics[width=1\linewidth]{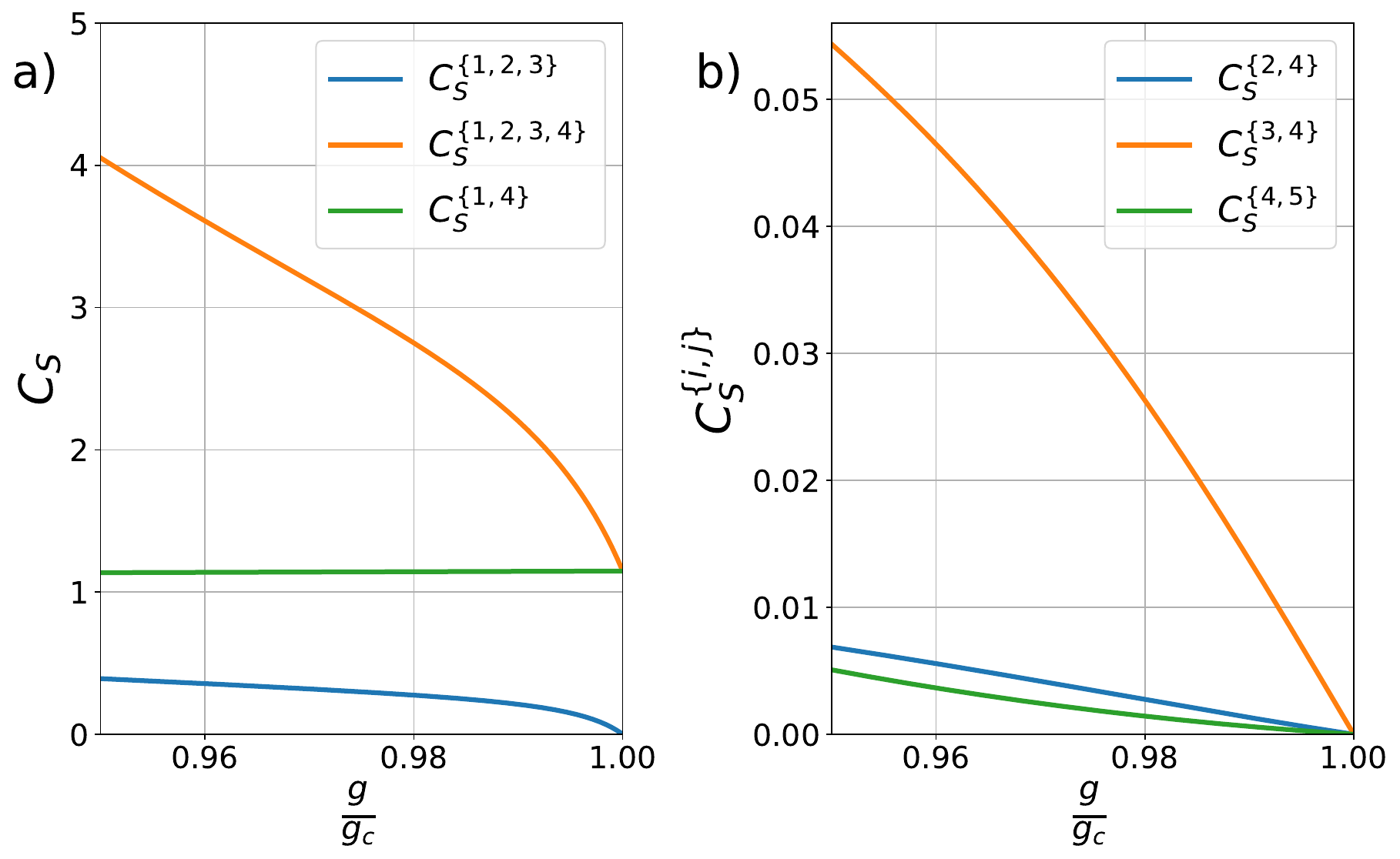}
    \caption{Plot of the scalar variance bound $C_S$ for various parameter subsets as a function of the normalized coupling $g/g_c$, employing the SS of the DD as a probe. Other parameters are fixed at $\omega_c = 1$, $\omega_a = 0.7$, $\kappa = 0.1$ and $\xi = 0.4$.}
    \label{fig:DD_ss_xi}
\end{figure}

First, we consider the scenario where the photon hopping $\xi$ is fixed and only the coupling $g$ is tuned towards criticality. As shown in Fig.~\ref{fig:DD_ss_xi}(a), this strategy does not yield a significant scaling advantage compared to the single-cavity scenario. Specifically, we find that at most three parameters can be estimated simultaneously with divergent precision. The scalar variance bound scales linearly, $C_S \sim (g_c-g)$, for both two- and three-parameter subsets not involving $\omega_c $ and $\xi$ simultaneously, mirroring the behavior of the DM. Instead, a particularly unfavorable case is the simultaneous estimation of $\omega_c$ and $\xi$: here, $C_S^{\{1,4\}}$ and also $C_S^{\{1,4, i\}}$ saturate to a finite value as $g \to g_c$.

To overcome these limitations, we again exploit TPs. Similarly to Section \ref{sec:DD_triple_point}, our strategy is to simultaneously tune $\xi$ and $g$ along a trajectory targeting the steady-state TP, located at:
\begin{equation}
    (\xi_t, g_t) = \left( 0, \frac{1}{2} \sqrt{ \omega_a \omega_c \left( 1 + \frac{\kappa^2}{\omega_c^2} \right) } \right).
\end{equation}
Again we choose a trajectory in the $\xi -g$ plane given by Eq.~\eqref{eq:traj}. Including dissipation, the value of $g_c(\xi)$ around $\xi = 0$ is modified to:
\begin{align}
    g_c(\xi) =& \notag \frac{1}{2} \sqrt{\omega_a \omega_c \left ( 1 + \frac{\kappa^2}{\omega_c^2} \right)} \\ &- \frac{1}{2} \frac{\omega_a | \omega_c^2- \kappa^2|}{ \sqrt{\omega_a \omega_c^3 ( \kappa^2+ \omega_c^2)}} | \xi| + O(\xi^2).
\end{align}
Consequently, to remain in the normal phase, the slope $k$ must satisfy:
\begin{equation}
    k < k_{\text{max}} = \frac{2 \sqrt{\omega_a \omega_c^3 ( \kappa^2+ \omega_c^2)}}{\omega_a | \omega_c^2- \kappa^2|}.
\end{equation}

\begin{figure*}
    \centering
    \includegraphics[width=0.9\linewidth]{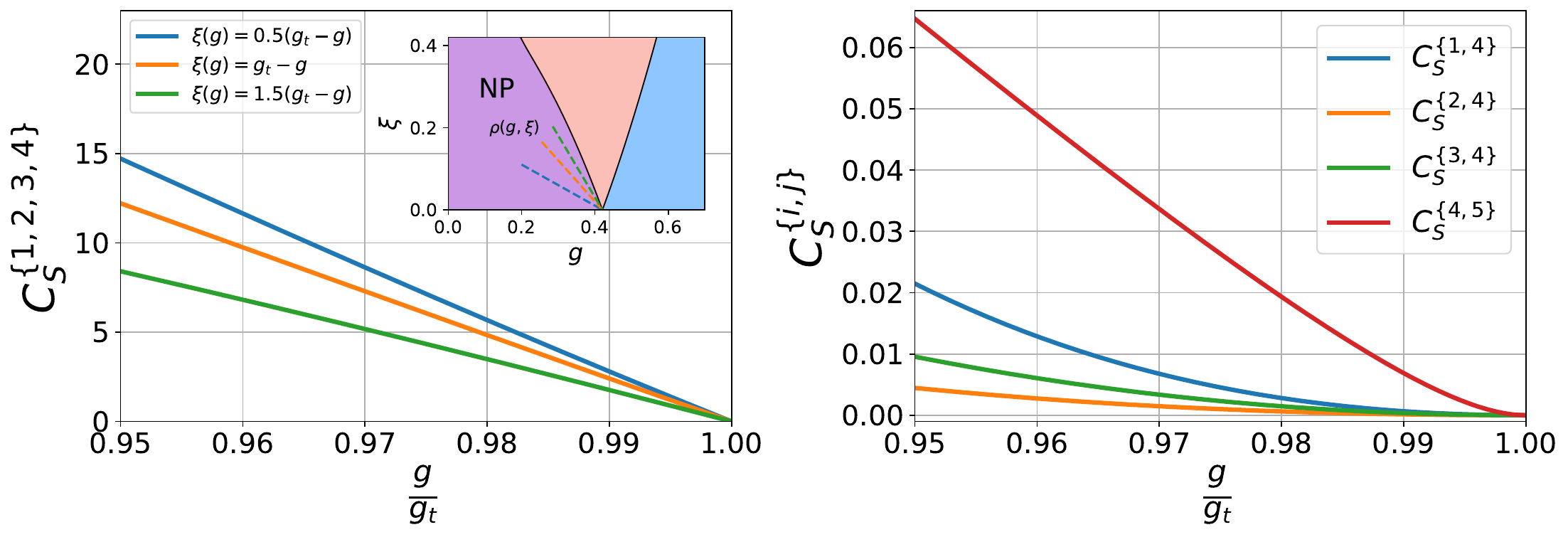}
    \caption{Plot of the scalar variance bound $C_S$ as a function of $g/g_t$, employing the SS of DD as a probe, where $\xi$ is tuned dynamically according to the trajectory Eq.~\eqref{eq:traj}. We fixed $\omega_c = 1, \omega_a = 0.7, \kappa = 0.1$. Panel (a) shows the effect of the trajectory slope $k$ on the precision in the simultaneous estimation of the parameters $\omega_c, g, \omega_a, \xi$, showing improvement as $k$ approaches $k_\text{max} \approx 2.43$. The inset represents the phase diagram of the DD in the $g-\xi$ plane, with the three different trajectories highlighted. Panel (b) shows the scaling of $C_S$ for the specific case $k = 1$, regarding the estimation of pairs of parameters involving $\xi$ and another parameter among $\omega_c, g, \omega_a$ or $\kappa$.}
    \label{fig:DD_ss_TP}
\end{figure*}

Tuning the system near a TP leads to two significant advantages. First, for two-parameter scenarios involving the hopping $\xi$ and another parameter, Fig.~\ref{fig:DD_ss_TP} demonstrates that we retrieve the quadratic scaling typical of single-parameter critical metrology:
\begin{equation}
    C_S^{\{i,4\}} \sim (g_t-g)^2.
\end{equation}
This confirms that the TP mechanism, simultaneously closing two gaps, restores the single-parameter scaling even in the dissipative SS.

Second, the simultaneous estimation of the four Hamiltonian parameters $\{\omega_c, g, \omega_a, \xi\}$ becomes possible with diverging precision. As shown in Fig.~\ref{fig:DD_ss_TP}, tuning along the triple-point trajectory yields a linear divergence of the precision:
\begin{equation}
    C_S^{\{1,2,3,4\}} \sim (g_t-g).
\end{equation}
This contrasts sharply with the fixed-$\xi$ case, where estimation was limited to finite precision. Furthermore, as can be seen from Fig. \ref{fig:DD_ss_TP}, a steeper approach trajectory (larger $k$) which lies closer to the critical line, improves the prefactor of the estimation precision.

Finally, when extending the analysis to the full set of five parameters (including the dissipation rate $\kappa$), we find that $\det[Q^{\{1,2,3,4,5\}}] \approx 0$ everywhere. Consequently, the simultaneous estimation of all five parameters remains impossible, even at finite precision.

\section{Comparison of the Results}
\label{sec:comparison}

In the previous sections, we comprehensively characterized the multi-parameter sensing capabilities of the DM and the DD, considering both the ground state and the steady state under photonic loss. We also demonstrated that tuning the system near a TP can yield a significant advantage in the DD scenario.

To effectively organize and compare these different strategies, we must consider the true operational resources required by the model. Specifically, we will assume the total time $T$ to be our fundamental resource.

\subsection{Analysis of the Resources}
Due to the phenomenon of critical slowing down, the time required to adiabatically tune the ground state to a specific value of the control parameter diverges as the system approaches the critical point. As detailed in Appendix \ref{app:time_cost}, the adiabatic preparation time $T_\text{A,DM}$ necessary to tune the ground state of the DM to a coupling $g$ near $g_c$ is given by:
\begin{equation}
    T_\text{A,DM} \approx \frac{2}{\gamma \Delta (g_c-g)^{1/2}}\,,
\end{equation}
where $\gamma$ is a small parameter governing the adiabatic speed and $\Delta = 4 \frac{(\omega_a \omega_c)^{3/4}}{\sqrt{\omega_a^2+ \omega_c^2}}$. Consequently, the adiabatic time diverges at the critical point as $T_\text{A,DM} \sim (g_c-g)^{-1/2}$. 

A similar reasoning applies to the DD. The adiabatic preparation time is found to be: 
\begin{equation}
    T_\text{A,DD} \approx \frac{2}{\gamma \Delta_1 (g_t-g)^{1/2}}\,,
\end{equation}
where $\Delta_1  = 2 \sqrt{2} \sqrt{\frac{ \omega_a \omega_c (2 \sqrt{\omega_a \omega_c} -k \omega_a)}{\omega_a^2+ \omega_c^2}}$.

In the dissipative scenario, the relevant temporal resource is instead the relaxation time, $T_\text{R}$, which is the time required for the system to reach its steady state. As shown in Appendix \ref{app:time_cost}, this time is determined by the lowest real part of the eigenvalues of the drift matrix. For these models, we found:
\begin{align}
   & T_\text{R,DM} \approx \frac{\kappa}{2 \sqrt{\frac{\omega_c \left(\kappa ^2+\omega_c^2\right)}{\omega_a}}} \frac{1}{g_c-g}\,,\\ &
     T_\text{R,DD} \approx \frac{\kappa}{2 \sqrt{\frac{\omega_c \left(\kappa ^2+\omega_c^2\right)}{\omega_a}}} \frac{1}{g_t-g}\,.
\end{align}
Notice that both relaxation times $ T_\text{R,DM} \sim (g_c -g)^{-1}$ and $ T_\text{R,DD} \sim (g_t -g)^{-1}$ diverge more rapidly than their adiabatic counterparts $ T_\text{A,DM} \sim (g_c -g)^{-1/2}$ and $ T_\text{A,DD} \sim (g_t -g)^{-1/2}$ .

\subsection{Trajectories in Parameter Space}
In Section \ref{sec:DD_steady_state_TP}, we suggested that optimizing the trajectory along which the system is tuned toward the TP can provide a metrological advantage. Specifically, when tuning the system along a linear trajectory defined by Eq.~\eqref{eq:traj}, increasing the slope $k$ improves the estimation precision for the steady state of the DD, as previously shown in Fig. \ref{fig:DD_ss_TP}. Crucially, this advantage persists even when accounting for time as a resource, because the relaxation time $T_\text{R,DD}$ is independent of $k$. 

\begin{figure}
    \centering
    \includegraphics[width=0.9\linewidth]{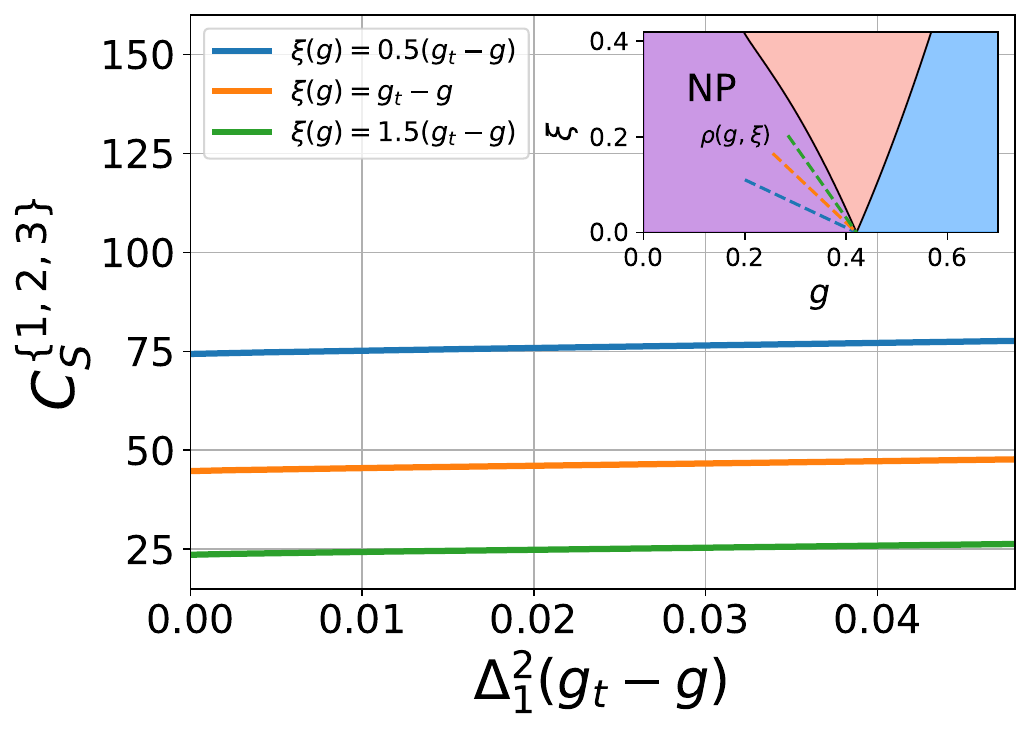}
    \caption{Plot of the scalar variance bound $C_S^{\{1,2,3\}}$ as a function of the temporally normalized parameter $\Delta_1^2 (g_t-g)$ for different approach trajectories for the GS of the DD. We chose slopes $k \in \{ 0.5, 1, 1.5\}$ and tuned the system along the linear trajectory described by Eq.~\eqref{eq:traj}. The inset represents the phase diagram of the DD in the $g-\xi$ plane, with the three different trajectories highlighted. The fixed system parameters are $\omega_c = 1$ and $\omega_a = 0.7$.}
    \label{fig:DD_k}
\end{figure}

However, this is no longer true when utilizing the ground state of the DD. The adiabatic preparation time $T_\text{A,DD}$ explicitly depends on $k$ through the parameter $\Delta_1$. Therefore, to rigorously verify whether an advantage exists, we must analyze the scalar variance bound $C_S$ as a function of the normalized temporal parameter $\Delta_1^2 (g_t-g)$. As illustrated in Fig. \ref{fig:DD_k}, an advantage is still observable within certain ranges of $k$. Notably, the limit $\Delta_1 \rightarrow 0$ as $k \rightarrow k_\text{max}$ reflects the physical reality that the adiabatic preparation time diverges if the trajectory runs too close to the critical phase boundary.

\subsection{Summary of the Results}
\label{sec:summary}

To synthesize the findings of the previous sections, we summarize the estimation precision scalings with respect to the fundamental resource time $T$ in Table \ref{tab:my_label}.

The table lists the asymptotic scaling of the scalar variance bound $C_S^{\{i_1, \dots, i_n\}}$ for all possible multiparameter combinations. We use a dash ($-$) to indicate that the system state does not depend on a specific combination of parameters. The symbol $S$ (for \textit{sloppy}) indicates that, while the state depends on all parameters in the subset, the determinant of the QFIM vanishes ($\det[Q^{\{i_1, \dots, i_n\}}] = 0$) everywhere, rendering simultaneous independent estimation impossible.

The abbreviations used in Table \ref{tab:my_label} are defined as follows: 
\begin{itemize}
    \item \textbf{GS DM:} Ground state of the Dicke model.
    \item \textbf{GS DD:} Ground state of the Dicke dimer with fixed photon hopping $\xi$.
    \item \textbf{GS DD (TP):} Ground state of the Dicke dimer where $g$ and $\xi$ are tuned simultaneously toward a triple point.
    \item \textbf{SS DM:} Steady state of the Dicke model under photon loss.
    \item \textbf{SS DD:} Steady state of the Dicke dimer with fixed $\xi$ under photon loss.
    \item \textbf{SS DD (TP):} Steady state of the Dicke dimer where $g$ and $\xi$ are tuned simultaneously toward a triple point under photon loss.
\end{itemize}

\begin{table*}[t]
\centering
\caption{Asymptotic scaling of the scalar variance bound $C_S$ as a function of the available resource time $T$ (for $T\rightarrow \infty$) across different models and parameter subsets. 'Par.s' denotes the parameters being estimated.}
\label{tab:my_label}
\begin{tabular}{|c c |c c c c c c |}
\hline
  \textbf{\# } & \textbf{Parameters} & \textbf{GS DM} & \textbf{GS DD} & \textbf{GS DD (TP)} & \textbf{SS DM} & \textbf{SS DD} & \textbf{SS DD (TP)} \\ [0.5ex]
  \hline
\multirow{10}{*}{2} &$\omega_c, g$ & $T^{-1}$ & $T^{-1}$ & $T^{-1}$ & $T^{-1}$ & $T^{-1}$ &  $T^{-1}$ \\
& $\omega_c, \omega_a$ & $T^{-1}$ & $T^{-1}$ & $T^{-1}$ & $T^{-1}$ & $T^{-1}$ &  $T^{-1}$ \\
&$\omega_c, \xi$ & - & $O(1)$ & $T^{-4}$ &-& $O(1)$ &  $T^{-2}$ \\
&$\omega_c, \kappa$ &-&-&-& $T^{-1}$  & $T^{-1}$ & $T^{-1}$ \\
& $\omega_a, g$& $T^{-1}$ & $T^{-1}$ & $T^{-1}$ & $T^{-1}$ & $T^{-1}$ &  $T^{-1}$ \\
&$\omega_a, \xi$& - & $T^{-1}$ & $T^{-4}$ & - & $T^{-1}$& $T^{-2}$ \\
&$\omega_a, \kappa$ &-&-&-& $T^{-1}$  & $T^{-1}$ & $T^{-1}$    \\
&$g, \xi$ & - & $T^{-1}$ & $T^{-4}$ & - & $T^{-1}$ &  $T^{-2}$ \\
&$g, \kappa$  &-&-&-& $T^{-1}$  & $T^{-1}$ & $T^{-1}$ \\
&$\xi, \kappa$  &-&-&-& -  & $T^{-1}$ & $T^{-2}$    \\
 \hline
 \multirow{10}{*}{3} & $\omega_c, g, \omega_a$ & S & $O(1)$ & $O(1)$ & $T^{-1}$ & $T^{-1}$ & $T^{-1}$ \\
&$\omega_c, g, \xi $ & - & $O(1)$ & $T^{-1}$ & - & $O(1)$ & $T^{-1}$ \\
&$\omega_c, g, \kappa $ &-&-&-&$T^{-1}$& $T^{-1}$ & $T^{-1}$ \\
&$\omega_c, \omega_a, \xi $&-& $O(1)$& $T^{-1}$&-& $O(1)$& $T^{-1}$ \\
&$\omega_c, \omega_a, \kappa $&-&-&-& $T^{-1}$ & $T^{-1}$& $T^{-1}$ \\
&$\omega_c, \xi, \kappa $ &-&-&-&-& $O(1)$ & $T^{-1}$ \\
&$\omega_a, g, \xi $&-&$O(1)$ & $T^{-1}$&-& $T^{-1}$ & $T^{-1}$ \\
&$\omega_a, g, \kappa $& -&-&-&$T^{-1}$ & $T^{-1}$& $T^{-1}$ \\
&$\omega_a, \xi, \kappa $&-& -&-&-&$T^{-1}$& $T^{-1}$ \\
&$g, \xi,\kappa$&-& -&-&-&$T^{-1}$& $T^{-1}$ \\
 \hline
 \multirow{5}{*}{4} & $\omega_c, g, \omega_a, \xi $ & - & S & S & - & $O(1)$ &$T^{-1}$\\
 & $\omega_c, g,\omega_a,\kappa $ & - & - & - & S & $O(1)$& $T^{-1}$ \\
 & $\omega_c, g,\xi,\kappa $ & - & - & - & - & $O(1)$& $T^{-1}$ \\
  & $\omega_c, \omega_a, \xi,\kappa $ & - & - & - & - & $O(1)$ &$T^{-1}$ \\
 & $ g, \omega_a,\xi, \kappa $ & - & - & - & - & $O(1)$ &$T^{-1}$ \\
 \hline
5 & $\omega_c, g, \omega_a, \xi, \kappa $&-&-&-&-&S&S\\
 \hline
\end{tabular}
\end{table*}

\section{Conclusions and Outlook}
\label{sec:conclusions}

In this work, we have systematically explored the potential of critical quantum metrology in a multiparameter setting, focusing on the paradigmatic DM and its lattice extension, the DD. Our central motivation was to overcome the phenomenon of \textit{sloppiness}, the rank deficiency of the QFIM that typically plagues critical systems, rendering the simultaneous independent estimation of multiple parameters impossible.

We have demonstrated that while the leading-order QFIM is often singular near a quantum phase transition, higher-order contributions can restore invertibility. For the single-cavity DM ground state, this allows for the simultaneous estimation of at most two parameters with a scalar variance bound scaling as $C_S \sim (g_c - g)^{1/2}$. Although this scaling is slower than the ideal single-parameter Heisenberg limit ($\sim (g_c-g)^2$), it represents a crucial proof of concept that multiparameter critical metrology is indeed feasible.

To recover the advantageous quadratic scaling, we introduced the DD with photon hopping. By tuning the system toward a TP, where two excitation gaps close simultaneously, we showed that the rank of the singular component of the QFIM increases. This enables the estimation of the photon hopping amplitude $\xi$ and another Hamiltonian parameter with the optimal quadratic scaling $C_S \sim (g_t - g)^2$. Furthermore, the TP unlocks the simultaneous estimation of up to three parameters with finite precision, a task that remains impossible in the single-cavity limit.

Furthermore, we addressed the practical challenge of dissipation. By analyzing the steady state (SS) under photonic loss, we found that critical enhancement is not washed out by environmental noise. Remarkably, in the dissipative DM, the simultaneous estimation of up to three parameters is possible with a scalar variance bound that vanishes linearly, $C_S \sim (g_c - g)$. Moreover, in the dissipative DD near the TP, specific pairs of parameters can be estimated with quadratically vanishing variance, $C_S \sim (g_t - g)^2$, and the simultaneous estimation of up to four parameters becomes possible with a linearly vanishing bound, $C_S \sim (g_t - g)$. This robustness suggests that critical metrology protocols can be implemented successfully in realistic open quantum systems without requiring perfect isolation.

Finally, we established a rigorous connection between the precision scaling near criticality and the fundamental resource of time, $T$. By explicitly calculating the adiabatic preparation time required for ground states and the relaxation time necessary to reach the steady state in lossy scenarios, we translated our parameter-dependent bounds into time-dependent bounds. This operational framework allowed us to fairly compare the different sensing strategies from a pure resource point of view.

Future research should investigate the extension of these results to larger Dicke lattices, where more complex critical points might allow for the simultaneous estimation of an even larger set of parameters. Additionally, exploring optimal control techniques to navigate the specific trajectories required to approach TPs adiabatically could further bridge the gap between theoretical bounds and experimental realization.

\section*{Acknowledgments}
LP thanks Francesco Albarelli for stimulating discussions and Arianna Magagna for the design of Fig. \ref{fig:rep_dicke}.

\bibliographystyle{unsrtnat}
\bibliography{bibliography}

@article{mondal_multicritical_2025,
  author        = {Mondal, Sayan and Sahoo, Ayan and Sen, Ujjwal and Rakshit, Debraj},
  title         = {Multicritical quantum sensors driven by symmetry-breaking},
  journal       = {Phys. Rev. B},
  volume        = {112},
  number        = {23},
  pages         = {235165},
  year          = {2025},
  doi           = {10.1103/PhysRevB.112.235165},
  eprint        = {2407.14428},
  archivePrefix = {arXiv},
  primaryClass  = {quant-ph},
  url           = {https://doi.org/10.1103/PhysRevB.112.235165}
}

@article{horodecki_five_2020,
  title   = {Five Open Problems in Quantum Information Theory},
  author  = {Horodecki, Pawe\l{} and Rudnicki, \L{}ukasz and \ifmmode \dot{Z}\else \.{Z}\fi{}yczkowski, Karol},
  journal = {PRX Quantum},
  volume  = {3},
  number  = {1},
  pages   = {010101},
  year    = {2022},
  month   = {Mar},
  doi     = {10.1103/PRXQuantum.3.010101},
  url     = {https://doi.org/10.1103/PRXQuantum.3.010101}
}

@misc{wei_boundary-induced_2025,
  author        = {Wei, Peng-Fei and Xu, Yilun and Sun, Fengxiao and He, Qiongyi and Rabl, Peter and Wang, Zhihai},
  title         = {Boundary-induced phases in the dissipative {Dicke} lattice model},
  year          = {2025},
  eprint        = {2508.10296},
  archivePrefix = {arXiv},
  primaryClass  = {quant-ph},
  doi           = {10.48550/arXiv.2508.10296},
  url           = {https://arxiv.org/abs/2508.10296}
}

@misc{mihailescu_critical_2025,
  author        = {Mihailescu, George and Alushi, Uesli and Di Candia, Roberto and Felicetti, Simone and Gietka, Karol},
  title         = {Critical quantum sensing: a tutorial on parameter estimation near quantum phase transitions},
  year          = {2025},
  eprint        = {2510.02035},
  archivePrefix = {arXiv},
  primaryClass  = {quant-ph},
  doi           = {10.48550/arXiv.2510.02035},
  url           = {https://arxiv.org/abs/2510.02035}
}

@article{hotter_quantum_2025,
  author        = {Hotter, Christoph and Miranowicz, Adam and Gietka, Karol},
  title         = {Quantum metrology in the ultrastrong coupling regime of light-matter interactions: Leveraging virtual excitations without extracting them},
  journal       = {Phys. Rev. Lett.},
  volume        = {135},
  number        = {10},
  pages         = {100802},
  year          = {2025},
  doi           = {10.1103/PhysRevLett.135.100802},
  eprint        = {2501.16304},
  archivePrefix = {arXiv},
  primaryClass  = {quant-ph},
  url           = {https://doi.org/10.1103/PhysRevLett.135.100802}
}

@article{mihailescu_uncertain_2025,
  author        = {Mihailescu, George and Campbell, Steve and Gietka, Karol},
  title         = {Uncertain quantum critical metrology: From single to multi-parameter sensing},
  journal       = {Phys. Rev. A},
  volume        = {111},
  number        = {5},
  pages         = {052621},
  year          = {2025},
  doi           = {10.1103/PhysRevA.111.052621},
  eprint        = {2407.19917},
  archivePrefix = {arXiv},
  primaryClass  = {quant-ph},
  url           = {https://doi.org/10.1103/PhysRevA.111.052621}
}

@article{garbe_critical_2020,
  author        = {Garbe, Louis and Bina, Matteo and Keller, Arne and Paris, Matteo G. A. and Felicetti, Simone},
  title         = {Critical quantum metrology with a finite-component quantum phase transition},
  journal       = {Phys. Rev. Lett.},
  volume        = {124},
  number        = {12},
  pages         = {120504},
  year          = {2020},
  doi           = {10.1103/PhysRevLett.124.120504},
  eprint        = {1910.00604},
  archivePrefix = {arXiv},
  primaryClass  = {quant-ph},
  url           = {https://doi.org/10.1103/PhysRevLett.124.120504}
}

@article{hotter_combining_2024,
  author        = {Hotter, Christoph and Ritsch, Helmut and Gietka, Karol},
  title         = {Combining critical and quantum metrology},
  journal       = {Phys. Rev. Lett.},
  volume        = {132},
  number        = {6},
  pages         = {060801},
  year          = {2024},
  doi           = {10.1103/PhysRevLett.132.060801},
  eprint        = {2311.16472},
  archivePrefix = {arXiv},
  primaryClass  = {quant-ph},
  url           = {https://doi.org/10.1103/PhysRevLett.132.060801}
}

@article{luo_quantum_2025,
  title         = {Quantum phase transitions in a Dicke trimer with both photon and atom hoppings},
  author        = {Luo, Jun-Wen and Wang, Bo and Xiang, Ze-Liang},
  journal       = {Phys. Rev. A},
  year          = {2026},
  note          = {Accepted},
  doi           = {10.1103/ywj2-hm54},
  eprint        = {2502.10839},
  archivePrefix = {arXiv},
  primaryClass  = {quant-ph},
  url           = {https://link.aps.org/doi/10.1103/ywj2-hm54}
}

@article{zhu_quantum_2024,
  author        = {Zhu, Xin and L{\"u}, Jia-Hao and Ning, Wen and Shen, Li-Tuo and Wu, Fan and Yang, Zhen-Biao},
  title         = {Quantum geometric tensor and critical metrology in the anisotropic {Dicke} model},
  journal       = {Phys. Rev. A},
  volume        = {109},
  number        = {5},
  pages         = {052621},
  year          = {2024},
  doi           = {10.1103/PhysRevA.109.052621},
  eprint        = {2406.06301},
  archivePrefix = {arXiv},
  primaryClass  = {quant-ph},
  url           = {https://doi.org/10.1103/PhysRevA.109.052621}
}

@article{gietka_understanding_2022,
  author        = {Gietka, Karol and Ruks, Lewis and Busch, Thomas},
  title         = {Understanding and improving critical metrology: Quenching superradiant light-matter systems beyond the critical point},
  journal       = {Quantum},
  volume        = {6},
  pages         = {700},
  year          = {2022},
  doi           = {10.22331/q-2022-04-27-700},
  eprint        = {2110.04048},
  archivePrefix = {arXiv},
  primaryClass  = {quant-ph},
  url           = {https://doi.org/10.22331/q-2022-04-27-700}
}

@article{xu_phase_2024,
  title     = {Phase Transition and Multistability in Dicke Dimer},
  author    = {Xu, Yilun and Sun, Feng-Xiao and Zhang, Wei and He, Qiongyi and Pu, Han},
  journal   = {Phys. Rev. Lett.},
  volume    = {133},
  number    = {23},
  pages     = {233604},
  year      = {2024},
  month     = {Dec},
  doi       = {10.1103/PhysRevLett.133.233604},
  url       = {https://doi.org/10.1103/PhysRevLett.133.233604},
  publisher = {American Physical Society}
}

@article{suzuki_quantum_2020,
  author        = {Suzuki, Jun and Yang, Yuxiang and Hayashi, Masahito},
  title         = {Quantum state estimation with nuisance parameters},
  journal       = {J. Phys. A: Math. Theor.},
  volume        = {53},
  number        = {45},
  pages         = {453001},
  year          = {2020},
  doi           = {10.1088/1751-8121/ab8b78},
  eprint        = {1911.02790},
  archivePrefix = {arXiv},
  primaryClass  = {quant-ph},
  url           = {https://doi.org/10.1088/1751-8121/ab8b78}
}

@article{zhou_quantum_2020,
  author  = {Zhou, Jian-Yong and Zhou, Yue-Hui and Yin, Xian-Li and Huang, Jin-Feng and Liao, Jie-Qiao},
  title   = {Quantum entanglement maintained by virtual excitations in an ultrastrongly coupled oscillator system},
  journal = {Sci. Rep.},
  volume  = {10},
  number  = {1},
  pages   = {12557},
  year    = {2020},
  doi     = {10.1038/s41598-020-68309-3},
  url     = {https://doi.org/10.1038/s41598-020-68309-3}
}

@article{albarelli2020upper,
  author        = {Tsang, Mankei and Albarelli, Francesco and Datta, Animesh},
  title         = {Quantum semiparametric estimation},
  journal       = {Phys. Rev. X},
  volume        = {10},
  number        = {3},
  pages         = {031023},
  year          = {2020},
  doi           = {10.1103/PhysRevX.10.031023},
  eprint        = {1906.09871},
  archivePrefix = {arXiv},
  primaryClass  = {quant-ph},
  url           = {https://doi.org/10.1103/PhysRevX.10.031023}
}

@article{suzuki_nuisance_2020,
  author        = {Suzuki, Jun},
  title         = {Nuisance parameter problem in quantum estimation theory: General formulation and qubit examples},
  journal       = {J. Phys. A: Math. Theor.},
  volume        = {53},
  number        = {26},
  pages         = {264001},
  year          = {2020},
  doi           = {10.1088/1751-8121/ab8672},
  eprint        = {1905.04733},
  archivePrefix = {arXiv},
  primaryClass  = {quant-ph},
  url           = {https://doi.org/10.1088/1751-8121/ab8672}
}

@article{mihailescu_multiparameter_2024,
  author  = {Mihailescu, George and Bayat, Abolfazl and Campbell, Steve and Mitchell, Andrew K.},
  title   = {Multiparameter critical quantum metrology with impurity probes},
  journal = {Quantum Sci. Technol.},
  volume  = {9},
  number  = {3},
  pages   = {035033},
  year    = {2024},
  doi     = {10.1088/2058-9565/ad438d},
  url     = {https://doi.org/10.1088/2058-9565/ad438d}
}

@misc{cheng_super-heisenberg_2025,
  author        = {Cheng, Jia-Ming and Zhang, Yong-Chang and Zhou, Xiang-Fa and Zhou, Zheng-Wei},
  title         = {Super-{Heisenberg} scaling in a triple point criticality},
  year          = {2025},
  eprint        = {2409.14048},
  archivePrefix = {arXiv},
  primaryClass  = {quant-ph},
  doi           = {10.48550/arXiv.2409.14048},
  url           = {https://arxiv.org/abs/2409.14048}
}

@article{genoni_conditional_2016,
  author        = {Genoni, Marco G. and Lami, Ludovico and Serafini, Alessio},
  title         = {Conditional and unconditional {Gaussian} quantum dynamics},
  journal       = {Contemp. Phys.},
  volume        = {57},
  number        = {3},
  pages         = {331--349},
  year          = {2016},
  doi           = {10.1080/00107514.2015.1125624},
  eprint        = {1607.02619},
  archivePrefix = {arXiv},
  primaryClass  = {quant-ph},
  url           = {https://doi.org/10.1080/00107514.2015.1125624}
}

@misc{chang_multiparameter_2025,
  author        = {Chang, Shoukang and Genoni, Marco G. and Albarelli, Francesco},
  title         = {Multiparameter quantum estimation with {Gaussian} states: Efficiently evaluating {Holevo}, {RLD} and {SLD} {Cram\'er-Rao} bounds},
  year          = {2025},
  eprint        = {2504.17873},
  archivePrefix = {arXiv},
  primaryClass  = {quant-ph},
  doi           = {10.48550/arXiv.2504.17873},
  url           = {https://arxiv.org/abs/2504.17873}
}

@article{he_scrambling_2025,
  title         = {Scrambling for precision: optimizing multiparameter qubit estimation in the face of sloppiness and incompatibility},
  author        = {He, Jiayu and Paris, Matteo G. A.},
  journal       = {J. Phys. A: Math. Theor.},
  volume        = {58},
  number        = {32},
  pages         = {325301},
  year          = {2025},
  publisher     = {IOP Publishing},
  doi           = {10.1088/1751-8121/adf585},
  eprint        = {2503.08235},
  archivePrefix = {arXiv},
  primaryClass  = {quant-ph},
  url           = {https://doi.org/10.1088/1751-8121/adf585}
}

@article{albarelli_perspective_2020,
  author        = {Albarelli, Francesco and Barbieri, Marco and Genoni, Marco G. and Gianani, Ilaria},
  title         = {A perspective on multiparameter quantum metrology: From theoretical tools to applications in quantum imaging},
  journal       = {Phys. Lett. A},
  volume        = {384},
  number        = {12},
  pages         = {126311},
  year          = {2020},
  doi           = {10.1016/j.physleta.2020.126311},
  eprint        = {1911.12067},
  archivePrefix = {arXiv},
  primaryClass  = {quant-ph},
  url           = {https://doi.org/10.1016/j.physleta.2020.126311}
}

@article{klinder_dynamical_2015,
  author        = {Klinder, J. and Ke{\ss}ler, H. and Wolke, M. and Mathey, L. and Hemmerich, A.},
  title         = {Dynamical phase transition in the open {Dicke} model},
  journal       = {Proc. Natl. Acad. Sci. U.S.A.},
  volume        = {112},
  number        = {11},
  pages         = {3290--3295},
  year          = {2015},
  doi           = {10.1073/pnas.1417132112},
  eprint        = {1409.1945},
  archivePrefix = {arXiv},
  primaryClass  = {cond-mat.quant-gas},
  url           = {https://doi.org/10.1073/pnas.1417132112}
}

@article{dicke_coherence_1954,
  author  = {Dicke, R. H.},
  title   = {Coherence in spontaneous radiation processes},
  journal = {Phys. Rev.},
  volume  = {93},
  number  = {1},
  pages   = {99--110},
  year    = {1954},
  doi     = {10.1103/PhysRev.93.99},
  url     = {https://doi.org/10.1103/PhysRev.93.99}
}

@article{baumann_dicke_2010,
  author  = {Baumann, Kristian and Guerlin, Christine and Brennecke, Ferdinand and Esslinger, Tilman},
  title   = {Dicke quantum phase transition with a superfluid gas in an optical cavity},
  journal = {Nature},
  volume  = {464},
  number  = {7293},
  pages   = {1301--1306},
  year    = {2010},
  doi     = {10.1038/nature09009},
  url     = {https://doi.org/10.1038/nature09009}
}

@article{giovannetti_advances_2011,
  author        = {Giovannetti, Vittorio and Lloyd, Seth and Maccone, Lorenzo},
  title         = {Advances in quantum metrology},
  journal       = {Nat. Photonics},
  volume        = {5},
  number        = {4},
  pages         = {222--229},
  year          = {2011},
  doi           = {10.1038/nphoton.2011.35},
  eprint        = {1102.2318},
  archivePrefix = {arXiv},
  primaryClass  = {quant-ph},
  url           = {https://doi.org/10.1038/nphoton.2011.35}
}

@article{zanardi_quantum_2008,
  author  = {Zanardi, Paolo and Paris, Matteo G. A. and Campos Venuti, Lorenzo},
  title   = {Quantum criticality as a resource for quantum estimation},
  journal = {Phys. Rev. A},
  volume  = {78},
  number  = {4},
  pages   = {042105},
  year    = {2008},
  doi     = {10.1103/PhysRevA.78.042105},
  url     = {https://doi.org/10.1103/PhysRevA.78.042105}
}

@article{liu_quantum_2020,
  author        = {Liu, Jing and Yuan, Haidong and Lu, Xiao-Ming and Wang, Xiaoguang},
  title         = {Quantum {Fisher} information matrix and multiparameter estimation},
  journal       = {J. Phys. A: Math. Theor.},
  volume        = {53},
  number        = {2},
  pages         = {023001},
  year          = {2020},
  doi           = {10.1088/1751-8121/ab5d4d},
  eprint        = {1907.08037},
  archivePrefix = {arXiv},
  primaryClass  = {quant-ph},
  url           = {https://doi.org/10.1088/1751-8121/ab5d4d}
}

@article{paris_quantum_2009,
  author  = {Paris, Matteo G. A.},
  title   = {Quantum estimation for quantum technology},
  journal = {Int. J. Quantum Inf.},
  volume  = {7},
  number  = {01},
  pages   = {125--137},
  year    = {2009},
  doi     = {10.1142/S0219749909004839},
  url     = {https://doi.org/10.1142/S0219749909004839}
}

@article{holstein_field_1940,
  author  = {Holstein, T. and Primakoff, H.},
  title   = {Field dependence of the intrinsic domain magnetization of a ferromagnet},
  journal = {Phys. Rev.},
  volume  = {58},
  number  = {12},
  pages   = {1098--1113},
  year    = {1940},
  doi     = {10.1103/PhysRev.58.1098},
  url     = {https://doi.org/10.1103/PhysRev.58.1098}
}

@book{serafini_quantum_2021,
  author    = {Serafini, Alessio},
  title     = {Quantum Continuous Variables: A Primer of Theoretical Methods},
  publisher = {CRC Press},
  address   = {Boca Raton},
  year      = {2021},
  isbn      = {9781482246346},
  url       = {https://isbnsearch.org/isbn/9781482246346}
}

@article{demkowicz-dobrzanski_multi-parameter_2020,
  author        = {Demkowicz-Dobrza{\'n}ski, Rafa{\l} and G{\'o}recki, Wojciech and Gu{\c t}{\u a}, M{\u a}d{\u a}lin},
  title         = {Multi-parameter estimation beyond quantum {Fisher} information},
  journal       = {J. Phys. A: Math. Theor.},
  volume        = {53},
  number        = {36},
  pages         = {363001},
  year          = {2020},
  doi           = {10.1088/1751-8121/ab8ef3},
  eprint        = {2001.11742},
  archivePrefix = {arXiv},
  primaryClass  = {quant-ph},
  url           = {https://doi.org/10.1088/1751-8121/ab8ef3}
}

@misc{pezze_advances_2025,
  author        = {Pezz{\`e}, Luca and Smerzi, Augusto},
  title         = {Advances in multiparameter quantum sensing and metrology},
  year          = {2025},
  eprint        = {2502.17396},
  archivePrefix = {arXiv},
  primaryClass  = {quant-ph},
  doi           = {10.48550/arXiv.2502.17396},
  url           = {https://arxiv.org/abs/2502.17396}
}

@misc{ferraro_gaussian_2005,
  author        = {Ferraro, Alessandro and Olivares, Stefano and Paris, Matteo G. A.},
  title         = {Gaussian states in continuous variable quantum information},
  year          = {2005},
  eprint        = {quant-ph/0503237},
  archivePrefix = {arXiv},
  primaryClass  = {quant-ph},
  doi           = {10.48550/arXiv.quant-ph/0503237},
  url           = {https://arxiv.org/abs/quant-ph/0503237}
}

@article{weedbrook_gaussian_2012,
  author        = {Weedbrook, Christian and Pirandola, Stefano and Garcia-Patron, Raul and Cerf, Nicolas J. and Ralph, Timothy C. and Shapiro, Jeffrey H. and Lloyd, Seth},
  title         = {Gaussian quantum information},
  journal       = {Rev. Mod. Phys.},
  volume        = {84},
  number        = {2},
  pages         = {621--669},
  year          = {2012},
  doi           = {10.1103/RevModPhys.84.621},
  eprint        = {1110.3234},
  archivePrefix = {arXiv},
  primaryClass  = {quant-ph},
  url           = {https://doi.org/10.1103/RevModPhys.84.621}
}

@book{cramer1999mathematical,
  author    = {Cram{\'e}r, Harald},
  title     = {Mathematical Methods of Statistics},
  volume    = {43},
  publisher = {Princeton University Press},
  year      = {1999},
  isbn      = {9780691005478},
  url       = {https://www.worldcat.org/isbn/9780691005478}
}

@book{Lehmann1998theor,
  author    = {Lehmann, E. L. and Casella, George},
  title     = {Theory of Point Estimation},
  edition   = {2nd},
  publisher = {Springer},
  series    = {Springer Texts in Statistics},
  year      = {1998},
  doi       = {10.1007/b98854},
  isbn      = {9780387985022},
  url       = {https://doi.org/10.1007/b98854}
}

@article{Helstrom1967,
  author  = {Helstrom, Carl W.},
  title   = {Minimum mean-squared error of estimates in quantum statistics},
  journal = {Phys. Lett. A},
  volume  = {25},
  number  = {2},
  pages   = {101--102},
  year    = {1967},
  doi     = {10.1016/0375-9601(67)90366-0},
  url     = {https://doi.org/10.1016/0375-9601(67)90366-0}
}

@article{carollo2019quantumness,
  author        = {Carollo, Angelo and Spagnolo, Bernardo and Dubkov, Alexander A. and Valenti, Davide},
  title         = {On quantumness in multi-parameter quantum estimation},
  journal       = {J. Stat. Mech. Theory Exp.},
  volume        = {2019},
  number        = {9},
  pages         = {094010},
  year          = {2019},
  doi           = {10.1088/1742-5468/ab3ccb},
  eprint        = {1911.11621},
  archivePrefix = {arXiv},
  primaryClass  = {quant-ph},
  url           = {https://doi.org/10.1088/1742-5468/ab3ccb}
}

@article{candeloro2021properties,
  author  = {Candeloro, Alessandro and Paris, Matteo G. A. and Genoni, Marco G.},
  title   = {On the properties of the asymptotic incompatibility measure in multiparameter quantum estimation},
  journal = {J. Phys. A: Math. Theor.},
  volume  = {54},
  number  = {48},
  pages   = {485301},
  year    = {2021},
  doi     = {10.1088/1751-8121/ac331e},
  url     = {https://doi.org/10.1088/1751-8121/ac331e}
}

@article{UHLMANN1986229,
  author  = {Uhlmann, Armin},
  title   = {Parallel transport and ``quantum holonomy'' along density operators},
  journal = {Rep. Math. Phys.},
  volume  = {24},
  number  = {2},
  pages   = {229--240},
  year    = {1986},
  doi     = {10.1016/0034-4877(86)90055-8},
  url     = {https://doi.org/10.1016/0034-4877(86)90055-8}
}

@article{montenegro2025quantum,
  title     = {Quantum metrology and sensing with many-body systems},
  author    = {Montenegro, Victor and Mukhopadhyay, Chiranjib and Yousefjani, Rozhin and Sarkar, Saubhik and Mishra, Utkarsh and Paris, Matteo G. A. and Bayat, Abolfazl},
  journal   = {Physics Reports},
  volume    = {1134},
  pages     = {1--62},
  year      = {2025},
  publisher = {Elsevier},
  doi       = {10.1016/j.physrep.2025.05.005},
  url       = {https://doi.org/10.1016/j.physrep.2025.05.005}
}

@article{zhao2025near,
  title={Near-Ultimate Quantum-Enhanced Sensitivity in Dissipative Critical Sensing with Partial Access},
  author={Zhao, Dingwei and Bayat, Abolfazl and Montenegro, Victor},
  journal={arXiv preprint arXiv:2508.19606},
  year={2025}
}

@article{fq4l-8v5g,
  title = {Quantum Fisher information as a witness of non-Markovianity and criticality in the spin-boson model},
  author = {Parlato, D. and Di Bello, G. and Pavan, F. and De Filippis, G. and Perroni, C. A.},
  journal = {Phys. Rev. B},
  volume = {112},
  issue = {22},
  pages = {224314},
  numpages = {10},
  year = {2025},
  month = {Dec},
  publisher = {American Physical Society},
  doi = {10.1103/fq4l-8v5g},
  url = {https://link.aps.org/doi/10.1103/fq4l-8v5g}
}

\onecolumn\newpage
\appendix

\section{Evaluation of the QFIM for the Ground States}
In this appendix we evaluate the QFIM associated with the ground state (GS) of the Dicke model (DM). The same strategy extends straightforwardly to the Dicke dimer (DD), as discussed in Sec.~\ref{app:a2}.

\subsection{QFIM for the DM}
\label{sec_qfim_dm}

To apply Eq.~\eqref{QFIpure}, it is enough to evaluate the overlap between the derivative of the GS and a basis contain the GS. Writing the GS as
\begin{equation}
    |G\rangle_{a,b} = \hat U \hat S_a \hat S_b |0,0\rangle_{a,b},
\end{equation}
we evaluate the overlaps of $\partial_\mu |G\rangle$ with eigenbasis of the DM Eq~\eqref{eq:eigenbasis_dm}:
\begin{align}
{}_{a,b}\langle \psi_{j,k} | \partial_\mu G \rangle_{a,b}
&=
{}_{a}\langle j|\, {}_{b}\langle k|
(\hat S_a \hat S_b)^\dagger
\hat U^\dagger (\partial_\mu \hat U)
\hat S_a \hat S_b
|0\rangle_a |0\rangle_b
\notag\\
&\quad+
{}_{a}\langle j|\, {}_{b}\langle k|
(\hat S_a \hat S_b)^\dagger
\partial_\mu(\hat S_a \hat S_b)
|0\rangle_a |0\rangle_b .
\end{align}

For a unitary operator $\hat V = e^{\hat O}$, one has
\begin{equation}
    \hat V^\dagger \partial_\mu \hat V
    =
    \sum_{n=0}^{\infty}
    \frac{(-1)^n}{(n+1)!}\,
    \hat O^{\times n}(\partial_\mu \hat O),
\end{equation}
where the nested commutators are defined recursively as
\begin{equation}
    \hat O^{\times 0}(\hat X)=\hat X,
    \qquad
    \hat O^{\times (n+1)}(\hat X)=[\hat O,\hat O^{\times n}(\hat X)].
\end{equation}

We now define
\begin{align}
    c_\pm &= \theta \frac{\omega_a\pm\omega_c}{\sqrt{\omega_c\omega_a}},
    \\
    \hat O_- &= \frac{1}{2}\left(\hat a^\dagger \hat b^\dagger-\hat a\hat b\right),
    \qquad
    \hat O_+ = \frac{1}{2}\left(\hat a \hat b^\dagger-\hat a^\dagger \hat b\right),
    \\
    \hat P_a &= \frac{1}{4}\left(\hat a^2-\hat a^{\dagger 2}\right),
    \qquad
    \hat P_b = \frac{1}{4}\left(\hat b^2-\hat b^{\dagger 2}\right).
\end{align}

These operators close under commutation according to
\begin{align}
    [\hat P_a-\hat P_b,\hat O_-] &= \hat O_+,
    \qquad
    [\hat P_a-\hat P_b,\hat O_+] = \hat O_-,
    \qquad
    [\hat O_+,\hat O_-] = \hat P_a-\hat P_b,
    \\
    [\hat P_a+\hat P_b,\hat O_-] &= 0,
    \qquad
    [\hat P_a+\hat P_b,\hat O_+] = 0,
    \qquad
    [\hat P_a,\hat P_b]=0.
\end{align}

Since $\hat U=e^{\hat O}$ with
\begin{equation}
    \hat O = c_- \hat O_- + c_+ \hat O_+,
\end{equation}
we obtain
\begin{align}
   \hat O^{\times(2j+1)}(\partial_\mu \hat O)
    &=
    \delta_\mu (-4\theta^2)^{j}
    (\hat P_a-\hat P_b),
    \qquad j\ge 0,
    \\
    \hat O^{\times(2j+2)}(\partial_\mu \hat O)
    &= -\delta_\mu (-4\theta^2)^j
    \left(c_+ \hat O_- + c_- \hat O_+\right),
    \qquad j\ge 0,    
\end{align}
where
\begin{equation}
    \delta_\mu = c_+\partial_\mu c_- - (\partial_\mu c_+)c_-.
\end{equation}
Using $c_+^2-c_-^2=4\theta^2$, the series can be resummed, yielding
\begin{align}
    \chi^\mu_{ab}
    &= -\frac{\delta_\mu \sin^2 \theta}{4 \theta^2},
    \\
    \chi^\mu_{\pm}
    &= \partial_\mu c_\pm
    - \frac{\delta_\mu c_\mp}{8\theta^3}
    \left(2\theta-\sin(2\theta)\right).
\end{align}
Therefore,
\begin{equation}
    \hat U^\dagger \partial_\mu \hat U
    =
    2\chi^\mu_{ab}(\hat P_a-\hat P_b)
    + \chi^\mu_{-}\hat O_-
    + \chi^\mu_{+}\hat O_+.
\end{equation}

We now evaluate the squeezing transformation of the $\hat O_\pm$ contribution. Defining
\begin{equation}
    \eta \equiv r_b-r_a,
    \qquad
    \hat X \equiv \eta(\hat P_a-\hat P_b),
    \qquad
    \hat Y_\mu \equiv \chi^\mu_{-}\hat O_-+\chi^\mu_{+}\hat O_+,
\end{equation}
we have
\begin{equation}
    (\hat S_a\hat S_b)^\dagger \hat Y_\mu \hat S_a\hat S_b
    =
    e^{\hat X}\hat Y_\mu e^{-\hat X}.
\end{equation}
Using BCH formula,
\begin{equation}
    e^{\hat X}\hat Y e^{-\hat X}
    =
    \sum_{n=0}^{\infty}\frac{1}{n!}\hat X^{\times n}(\hat Y),
\end{equation}
with
\begin{equation}
    \hat X^{\times 0}(\hat Y)=\hat Y,
    \qquad
    \hat X^{\times (n+1)}(\hat Y)=[\hat X,\hat X^{\times n}(\hat Y)],
\end{equation}
one finds
\begin{align}
    \hat X^{\times (2j)}(\hat Y_\mu)
    &=
    \eta^{2j}\hat Y_\mu,
    \\
    \hat X^{\times (2j+1)}(\hat Y_\mu)
    &=
    \eta^{2j+1}
    \left(\chi^\mu_{-}\hat O_+ + \chi^\mu_{+}\hat O_-\right).
\end{align}
Consequently,
\begin{align}
    (\hat S_a\hat S_b)^\dagger \hat Y_\mu \hat S_a\hat S_b
    &=
    \cosh(\eta)\left(\chi^\mu_{-}\hat O_-+\chi^\mu_{+}\hat O_+\right)
    \notag\\
    &\quad+
    \sinh(\eta)\left(\chi^\mu_{+}\hat O_-+\chi^\mu_{-}\hat O_+\right).
\end{align}

Moreover,
\begin{equation}
    (\hat S_a\hat S_b)^\dagger \partial_\mu(\hat S_a\hat S_b)
    =
    2(\partial_\mu r_a)\hat P_a + 2(\partial_\mu r_b)\hat P_b.
\end{equation}

Collecting all contributions, the derivative of the ground state in the $(a,b)$ Fock basis reads
\begin{align}
    {}_{a,b}\langle \psi_{j,k} |\partial_\mu G\rangle_{a,b}
    =
    \frac{1}{2}  {}_{a,b}\langle j,k |
    \Big[
    &-\left(\partial_\mu r_a+\chi^\mu_{ab}\right)\hat a^{\dagger 2}
    -\left(\partial_\mu r_b-\chi^\mu_{ab}\right)\hat b^{\dagger 2}
    \notag\\
    &\quad+
    \left(
    \cosh(r_b-r_a)\chi^\mu_{-}
    +\sinh(r_b-r_a)\chi^\mu_{+}
    \right)\hat a^\dagger \hat b^\dagger
    \Big]|0,0\rangle_{a,b}.
\end{align}

Therefore, the only non-zero overlaps with excited states are
\begin{align}
    {}_{a,b}\langle \psi_{2,0}|\partial_\mu G\rangle_{a,b}
    &=
    -\frac{1}{\sqrt{2}}
    \left(\partial_\mu r_a+\chi^\mu_{ab}\right),
    \\
    {}_{a,b}\langle \psi_{0,2}|\partial_\mu G\rangle_{a,b}
    &=
    -\frac{1}{\sqrt{2}}
    \left(\partial_\mu r_b-\chi^\mu_{ab}\right),
    \\
    {}_{a,b}\langle \psi_{1,1}|\partial_\mu G\rangle_{a,b}
    &=
    \frac{1}{2}
    \left(
    \cosh(r_b-r_a)\chi^\mu_{-}
    +\sinh(r_b-r_a)\chi^\mu_{+}
    \right).
\end{align}

The QFIM of the total system then follows from Eq.~\eqref{QFIpure}:
\begin{align}
    Q_{\mu\nu}
    &=
    2\left(\partial_\mu r_a+\chi^\mu_{ab}\right)\left(\partial_\nu r_a+\chi^\nu_{ab}\right)
    +
    2\left(\partial_\mu r_b-\chi^\mu_{ab}\right)\left(\partial_\nu r_b-\chi^\nu_{ab}\right)
    \notag\\
    &\quad+
    \left[
    \cosh(r_b-r_a)\chi^\mu_{-}
    +\sinh(r_b-r_a)\chi^\mu_{+}
    \right]
    \left[
    \cosh(r_b-r_a)\chi^\nu_{-}
    +\sinh(r_b-r_a)\chi^\nu_{+}
    \right].
\end{align}

\subsection{QFIM for the DD}
\label{app:a2}

For the DD, the derivation is completely analogous. Using the ground-state expression in Eq.~\eqref{eq:gs_dd}, the state factorizes into two independent Gaussian mode pairs, so the pure-state QFIM is additive over $n\in\{1,2\}$. One obtains
\begin{equation}
\begin{split}
    Q_{\mu\nu}
    &=
    \sum_{n=1}^{2}
    \Big[
    2(\partial_\mu r_{A_n}+\chi^\mu_{ab,n})(\partial_\nu r_{A_n}+\chi^\nu_{ab,n})
    +
    2(\partial_\mu r_{B_n}-\chi^\mu_{ab,n})(\partial_\nu r_{B_n}-\chi^\nu_{ab,n})
    \\
    &\qquad\qquad+
    \left(
    \cosh(r_{B_n}-r_{A_n})\chi^\mu_{-,n}
    +
    \sinh(r_{B_n}-r_{A_n})\chi^\mu_{+,n}
    \right)
    \\
    &\qquad\qquad\quad\times
    \left(
    \cosh(r_{B_n}-r_{A_n})\chi^\nu_{-,n}
    +
    \sinh(r_{B_n}-r_{A_n})\chi^\nu_{+,n}
    \right)
    \Big],
\end{split}
\label{eq:QFIM_DD_app}
\end{equation}
where, for each mode $n\in\{1,2\}$, we define
\begin{align}
    c_{\pm,n}
    &= \theta_n
    \frac{\omega_a \pm \overline{\omega}_{c,n}}
    {\sqrt{\overline{\omega}_{c,n}\omega_a}},
    \qquad
    \delta_{\mu,n}
    = c_{+,n}\partial_\mu c_{-,n}-(\partial_\mu c_{+,n})c_{-,n},
    \\
    \chi^\mu_{ab,n}
    &=
    -\frac{\delta_{\mu,n} \sin^2\theta}{4\theta_n^2},
    \\
    \chi^\mu_{\pm,n}
    &=
    \partial_\mu c_{\pm,n}-\frac{\delta_{\mu,n}c_{\mp,n}}{8\theta_n^3}
    \left(2\theta_n-\sin(2\theta_n)\right).
\end{align}

\section{Evaluation of the Time Costs}
\label{app:time_cost}
This appendix is dedicated to explicitly evaluating the time resource $T$ necessary to perform the estimation strategies described in the main text. In particular, we will evaluate the time necessary to adiabatically tune both the DM and the DD to their critical points in the noiseless scenario. Regarding the lossy scenario, we will evaluate the relaxation time necessary to reach the steady state.

\subsection{Adiabatic Time the DM}
In order to ensure the tuning process is adiabatic, we must impose the condition that, when tuning the system toward a critical point along a trajectory $g = g(t)$, the probability of observing an excitation remains small \cite{garbe_critical_2020,cheng_super-heisenberg_2025}. Specifically, for the DM, the state at time $t$ can be decomposed over the eigenbasis given in Eq.~\eqref{eq:eigenbasis_dm}:
\begin{equation}
   | \phi(t) \rangle_{a,b} = \sum_{m,n} c_{m,n}(t) e^{-i \theta_{m,n}(t)} |\psi_{m,n}(t) \rangle_{a,b}\,,
\end{equation}
where $\theta_{m,n} = \int_0^t (m \omega_+(t^\prime)+ n \omega_-(t^\prime) + \omega_0(t^\prime) )dt'$ and $\omega_0(t) $ is the GS energy. 

Assuming the initial state is the GS, at $t = 0$ we have the initial conditions $c_{0,0}(t = 0) = 1$ and $c_{m,n}(t = 0) = 0$ for $m,n \neq 0$. We can compute the evolution using the Schr\"odinger equation:
\begin{equation}
    i \frac{\partial}{\partial t} | \phi(t) \rangle = H(t) | \phi(t) \rangle\,,
\end{equation}
which leads to:
\begin{equation}
    \frac{\partial}{\partial t} c_{m,n}(t) = -\sum_{j,k} c_{j,k}(t) e^{i(\theta_{m,n}(t)- \theta_{j,k}(t))} {}_{a,b} \langle \psi_{m,n}(t) | \frac{\partial}{ \partial t} \psi_{j,k} (t)\rangle_{a,b}\,.
\end{equation}
According to time-dependent perturbation theory, we obtain:
\begin{align}\label{eq:cmn_dm}
    c_{m,n}(t) &= - \int_0^t e^{i[\theta_{m,n}(t^\prime)-\theta_{0,0}(t^\prime)]}{}_{a,b}\langle \psi_{m,n}(t^\prime)| \frac{\partial}{\partial t^\prime  }\psi_{0,0} (t^\prime)\rangle_{a,b} dt^\prime \\& \approx \int_0^t e^{i[\theta_{m,n}(t^\prime)-\theta_{0,0}(t^\prime)]} \frac{1}{\sqrt{2}} \delta_{n,2}  \delta_{m,0}  \partial_{t^{\prime}} r_b(t^\prime) dt^\prime\,. \notag
\end{align}
The adiabaticity condition then requires:
    \begin{equation}
        |c_{m,n}(t)|^2 \ll 1 \quad \text{for } m,n \neq 0\,.
    \end{equation}
From Eq.~\eqref{eq:cmn_dm}, it is clear that the only relevant excitation is given by:
\begin{align}
    c_{0,2} & \approx \int_0^{t} \exp\left(i \int_0^{t'} 2\omega_-(t^{\prime \prime}) d t^{\prime \prime } \right)\frac{1}{2 \sqrt{2}} \frac{\partial_{t'} \omega_-(t')}{ \omega_-(t')}dt' \notag \\ & = \int_0^{g} e^{i R(g' )}f(g') dg'\,,
\end{align}
where $R(g) = \int_0^{g} \frac{2\omega_-(g^{\prime})}{v(g^{\prime })} d g^{\prime  }$ and $f (g ) = \frac{1}{2 \sqrt{2}} \frac{\partial_{g} \omega_-(g)}{ \omega_-(g)}$, having set $dg = v(t) dt$. In order for the integral to be small, we require a tuning speed $v(g) $ such that:
\begin{equation}
   \left | \frac{\partial_g f(g)}{f(g)} \right | \ll \left | \partial_g R(g)\right|\,.
\end{equation}
Since:
\begin{equation}
    \left | \frac{\partial_g f(g)}{f(g)} \right | \approx \frac{1}{g_c-g}, \quad \text{and} \quad | \partial_g R(g)| \approx \Delta \frac{\sqrt{g_c-g}}{v(g)}\,,
\end{equation}
where $\Delta = 4 \frac{(\omega_a \omega_c)^{3/4}}{\sqrt{\omega_a^2+ \omega_c^2}} $, we find:
\begin{equation}
    v(g) = \gamma \Delta (g_c-g)^{3/2}\,,
\end{equation}
implying that the adiabatic preparation time is:
\begin{equation}
    T_\text{A,DM} = \int_0^g \frac{dg'}{v(g')} \approx \frac{2}{\gamma \Delta (g_c-g)^{1/2}}\,.
\end{equation}

\subsection{Adiabatic Time for the DD Tuned to a TP}
The situation is mathematically similar for the DD, for which we have:
\begin{align}
    &c_{m,n,p,q}(t)  \notag = - \int_0^t e^{i[\theta_{m,n,p,q}(t^\prime)-\theta_0(t^\prime)]}\langle \psi_{m,n,p,q}(t^\prime)| \frac{\partial}{\partial t^\prime  }\psi_{0}(t^\prime)\rangle dt^\prime \\ & \; \approx \int_0^t e^{i[\theta_{m,n,p,q}(t^\prime)-\theta_0(t^\prime)]} \frac{1}{\sqrt{2}} \big(  \delta_{m,0}\delta_{n,2} \delta_{p,0} \delta_{q,0} \partial_{t^{\prime}}  r_{B_1}(t^\prime)+\delta_{m,0}\delta_{n,0} \delta_{p,0} \delta_{q,2} \partial_{t^{\prime}} r_{B_2}(t^\prime)\big)  dt^\prime\,, 
\end{align}
where, for simplicity, we define $ |\psi_{m,n,p,q}(t) \rangle = |\psi_{m,n} (t)\rangle_{A_1, B_1} |\psi_{p,q}(t) \rangle_{A_2, B_2}$ and $ |\psi_0(t) \rangle = | \psi_{0,0,0,0}(t) \rangle $. Moreover, the dynamical phases are $\theta_{m,n,p,q}(t) = \int_0^t dt' ( m \omega_{+,1}(t^\prime)+n \omega_{-,1}(t^\prime)+ p \omega_{+,2}(t^\prime)+ q \omega_{-,2}(t^\prime)+ \omega_0(t))$ and $\theta_0 = \theta_{(0,0,0,0)} (t)$, with $\omega_0$ being the GS energy. 

Defining $c_1 = c_{0,2,0,0}$ and $c_2 =  c_{0,0,0,2} $, and assuming the tuning trajectory $g = g_t- \varepsilon$, $\xi = k \varepsilon$, with $d\varepsilon = -v(t) dt $, we obtain:
\begin{align}
    c_j(t) & \approx \int_0^{t} \exp\left(i \int_0^{t'} 2\omega_{-,j}(t^{\prime \prime}) d t^{\prime \prime } \right)\frac{1}{2 \sqrt{2}} \frac{\partial_{t'} \omega_{-,j}(t')}{ \omega_{-,j}(t')}dt' \notag \\ & = \int_0^{\varepsilon} e^{i R_j(\varepsilon' )}f_j(\varepsilon') d\varepsilon'\,,
\end{align}
where $R_j(\varepsilon) = \int_0^{\varepsilon} \frac{2\omega_{-,j}(\varepsilon^{\prime})}{v(\varepsilon^{\prime })} d \varepsilon^{\prime  }$ and $f_j (\varepsilon ) = \frac{1}{2 \sqrt{2}} \frac{\partial_{\varepsilon} \omega_{-,j}(\varepsilon)}{ \omega_{-,j}(\varepsilon)}$. In order to ensure $| c_1(t)|^2, | c_2(t)|^2 \ll 1 $, we require a velocity $v(\varepsilon) $ such that \cite{garbe_critical_2020,cheng_super-heisenberg_2025}:
\begin{equation}
   \left | \frac{\partial_\varepsilon f(\varepsilon)}{f(\varepsilon)} \right | \ll \left | \partial_\varepsilon R(\varepsilon)\right|\,.
\end{equation}
Since:
\begin{equation}
    \left | \frac{\partial_\varepsilon f_j (\varepsilon)}{f_j(\varepsilon)} \right |  \approx \frac{1}{\varepsilon} \quad \text{and} \quad | \partial_\varepsilon R_j(\varepsilon)| \approx \Delta_{j} \frac{\sqrt{\varepsilon}}{v(\varepsilon)}\,,
\end{equation}
where $\Delta_j  = 2 \sqrt{2} \sqrt{\frac{ \omega_a \omega_c (2 \sqrt{\omega_a \omega_c} + (-1)^j k \omega_a)}{\omega_a^2+ \omega_c^2}} $, being $\Delta_1 \le \Delta_2$, we find:
\begin{equation}
    v(g) = \gamma \Delta_1 (g_t-g)^{3/2}\,,
\end{equation}
implying the adiabatic preparation time for the DD is:
\begin{equation}
    T_\text{A,DD} = \int_0^{g_t} \frac{dg'}{v(g')} \approx \frac{2}{\gamma \Delta_1 (g_t-g)^{1/2}}\,.
\end{equation}

\subsection{Steady State Time for the DM}
The time necessary to reach the steady state (SS) is given by the inverse of the smallest absolute value of the real part of the eigenvalues of the drift matrix \cite{garbe_critical_2020}. Specifically for the DM, the drift matrix $A_\text{DM}$ from Eq.~\eqref{eq:drift_matrix_gs} can be expanded as:
\begin{equation}
    A_\text{DM} = A_{\text{DM},0}+ (g_c-g) B_\text{DM}\,,
\end{equation}
where $A_{\text{DM},0} = A_\text{DM}|_{g= g_c}$ and
\begin{equation}
    B_\text{DM} = \left(
\begin{array}{cccc}
 0 & 0 & 0 & 0 \\
 0 & 0 & 2 & 0 \\
 0 & 0 & 0 & 0 \\
 2 & 0 & 0 & 0 \\
\end{array}
\right)\,.
\end{equation}
Because the null space of $A_{\text{DM},0} $ and $A_{\text{DM},0}^T$ is one-dimensional and all other eigenvalues possess non-null real parts, we define $\text{Ker} [A_{\text{DM},0}] = \text{span} \{ v_0\}$ and $\text{Ker} [A_{\text{DM},0}^T] = \text{span} \{ u_0\}$, with the vectors:
\begin{align}
    v_0 &= \left\{-\sqrt{\frac{\omega_a \omega_c}{\kappa ^2+\omega_c^2}},-\frac{\kappa  \omega_a}{\sqrt{\omega_a \omega_c \left(\kappa ^2+\omega_c^2\right)}},1,0\right\}^T\,, \\
    u_0 &= \left\{-\frac{\kappa  \omega_a}{\sqrt{\omega_a \omega_c \left(\kappa ^2+\omega_c^2\right)}},-\sqrt{\frac{\omega_a \omega_c}{\kappa
   ^2+\omega_c^2}},0,1\right\}^T\,.
\end{align}
Using perturbation theory, the smallest real part of the eigenvalues $\lambda_j$ of $A_\text{DM}$ is:
\begin{equation}
    \min|\text{Re}{\lambda_j}| = \frac{u_0^T B v_0}{u_0^T v_0} (g_c-g) + \mathcal{O}( (g_c-g)^2)  = \frac{2 \sqrt{\frac{\omega_c \left(\kappa ^2+\omega_c^2\right)}{\omega_a}}}{\kappa } (g_c-g) + \mathcal{O}( (g_c-g)^2)\,.
\end{equation}
Consequently, the relaxation time reads:
\begin{equation}
    T_\text{R,DM} \approx \frac{\kappa}{2 \sqrt{\frac{\omega_c \left(\kappa ^2+\omega_c^2\right)}{\omega_a}}} \frac{1}{g_c-g}\,.
\end{equation}

\subsection{Steady State Time for the DD Tuned to a TP}
Similar to the DM, we expand the DD drift matrix from Eq.~\eqref{eq:drift_matrix_ss} to first order in the approach parameter $\varepsilon$:
\begin{equation}
    A_\text{DD} = A_{\text{DD},0} + \varepsilon B_\text{DD}+ \mathcal{O}( \varepsilon^2)\,.
\end{equation}
The left and right kernels of $A_{\text{DD},0}  = A_\text{DD}|_{(g,\xi) = (g_t,\xi_t)}$ are now degenerate. Specifically, $\text{Ker}[A_{\text{DD},0} ] = \text{span} \{ v_0, v_1\}$ and $\text{Ker}[A_{\text{DD},0}^T] = \text{span} \{ u_0, u_1\}$. We organize these vectors into matrices $V = ( v_0,v_1)$ and $U = (u_0, u_1)$. To simplify the notation, we define the auxiliary variables:
\begin{align}
    x = -\frac{\sqrt{\frac{\omega_c}{\kappa }}}{\sqrt{2}}\,, \quad 
    y = -\frac{\kappa }{\sqrt{2} \sqrt{\kappa \omega_c}}\,, \quad 
    z = \frac{1}{\sqrt{2} \sqrt{\frac{\kappa  \omega_a}{\kappa ^2+\omega_c^2}}}\,.
\end{align}
Using these variables, the kernel matrices are:
\begin{align}
    V &= \left(
\begin{array}{cccccccc}
 0 & 0 & 0 & 0 & x & y & z & 0 \\
 x & y & z & 0 & 0 & 0 & 0 & 0 \\
\end{array}
\right)^T\,, \\
    U &= \left(
\begin{array}{cccccccc}
 0 & 0 & 0 & 0 & y & x & 0 & z \\
 y & x & 0 & z & 0 & 0 & 0 & 0 \\
\end{array}
\right)^T\,.
\end{align}
We then construct the perturbation matrix $C$:
\begin{equation}
    C = U^T B_\text{DD} V\,.
\end{equation}
The eigenvalues of $C$, $\lambda_1, \lambda_2$, provide the first-order correction to the null eigenvalues of $A_0$, yielding:
\begin{equation}
   |\lambda_1| = \frac{2 \sqrt{\frac{\omega_c \left(\kappa ^2+\omega_c^2\right)}{\omega_a}}}{\kappa }\,, \quad  |\lambda_2| = \frac{2 \sqrt{\frac{\omega_c \left(\kappa ^2+\omega_c^2\right)}{\omega_a}}}{\kappa }+k
   \left(\frac{2 \omega_c}{\kappa }-\frac{2 \kappa }{\omega_c}\right)\,.
\end{equation}
For the parameter regime $\kappa < \omega_c $ and $k < k_\text{max}$, we have $| \lambda_1 | < | \lambda_2 |$. Therefore, the relaxation time is determined by the smallest eigenvalue correction, resulting in: 
\begin{equation}
    T_\text{R,DD} \approx \frac{\kappa}{2 \sqrt{\frac{\omega_c \left(\kappa ^2+\omega_c^2\right)}{\omega_a}}} \frac{1}{g_t-g}\,,
\end{equation}
which, notably, does not depend on the trajectory slope $k$.
\end{document}